\documentclass[fleqn,usenatbib]{mnras}

\usepackage{mathptmx}

\usepackage[T1]{fontenc}

\usepackage{amsmath}	
\usepackage{graphicx}	
\usepackage{amssymb}	
\usepackage{subfig}
\usepackage{multirow}
\usepackage{txfonts}
\usepackage{xcolor}

\title[A \textit{Chandra} study of Abell 795]{A \textit{Chandra} study of Abell 795 - a sloshing cluster with a FR0 radio galaxy at its center}

\author[F. Ubertosi et al.]{
F. Ubertosi$^{1,2}$\thanks{E-mail: francesco.ubertosi2@unibo.it},
M. Gitti$^{1,3}$,
E. Torresi$^{2}$,
F. Brighenti$^{1,4}$
and P. Grandi$^{2}$
\\
$^{1}$Dipartimento di Fisica e Astronomia (DIFA), Università di Bologna, via Gobetti 93/2, I-40129 Bologna, Italy\\
$^{2}$Istituto Nazionale di Astrofisica (INAF) - Osservatorio di Astrofisica e Scienza dello Spazio (OAS), via Gobetti 101, I-40129 Bologna, Italy\\
$^{3}$Istituto Nazionale di Astrofisica (INAF) - Istituto di Radioastronomia (IRA), via Gobetti 101, I-40129 Bologna, Italy\\
$^{4}$University of California Observatories/Lick Observatory, Department of Astronomy and Astrophysics, University of California, Santa Cruz, CA 95064, USA
}

\date{Accepted 2021 March 15. Received 2021 February 18; in original form 2020 December 29}
\pubyear{2021}

\begin{document}
\label{firstpage}
\pagerange{\pageref{firstpage}--\pageref{lastpage}}
\maketitle

\begin{abstract}
We present the first X-ray dedicated study of the galaxy cluster A795 and of the Fanaroff-Riley Type 0 hosted in its brightest cluster galaxy. Using an archival 30 ks \textit{Chandra} observation we study the dynamical state and cooling properties of the intracluster medium, and we investigate whether the growth of the radio galaxy is prevented by the surrounding environment. We discover that A795 is a weakly cool core cluster, with an observed mass deposition rate $\lessapprox 14\,$ M$_{\odot}$yr$^{-1}$ in the cooling region (central $\sim$66 kpc). In the inner $\sim$ 30 kpc we identify two putative X-ray cavities, and we unveil the presence of two prominent cold fronts at $\sim$60 kpc and $\sim$178 kpc from the center, located along a cold ICM spiral feature. The central galaxy, which is offset by 17.7 kpc from the X-ray peak, is surrounded by a multi-temperature gas with an average density of $n_{\text{e}} = 2.14 \times 10^{-2}$ cm$^{-3}$. We find extended radio emission at 74-227 MHz centered on the cluster, exceeding the expected flux from the radio galaxy extrapolated at low frequency. We propose that sloshing is responsible for the spiral morphology of the gas and the formation of the cold fronts, and that the environment alone cannot explain the compactness of the radio galaxy. We argue that the power of the two cavities and the sloshing kinetic energy can reduce and offset cooling. Considering the spectral and morphological properties of the extended radio emission, we classify it as a candidate radio mini-halo.
\end{abstract}




\section{Introduction}
Before the advent of high-resolution X-ray telescopes, the discovery of gas surface brightness peaks at the center of galaxy clusters led to the formulation of the standard cooling flow model, which predicted the occurrence of a pressure-driven inflow of cold gas in the core, at rates of up to 1000 M$_{\odot}$yr$^{-1}$ \citep{1994ARA&A..32..277F}. However, multi-wavelength observations probed that the amount of intracluster medium (ICM) that is actually cooling and flowing to the center is lower than theoretical predictions (e.g., \citealt{2001MNRAS.328..762E,2001A&A...365L.104P,2006PhR...427....1P}); moreover, high resolution X-ray spectra obtained with \textit{XMM-Newton} revealed the lack of soft emission expected from the gas cooling below 1-2 keV (e.g., \citealt{2001ApJ...560..194M,2002MNRAS.332L..50F,2002A&A...382..804B}).
\\Albeit reduced, cooling is observed in the so called cool core clusters (CCs),  which show a clear temperature drop in the central regions and the presence of cold gas in the innermost kpc, whereas it becomes negligible in non-cool core clusters (NCCs), typically associated to powerful merger events (e.g., \citealt{2006MNRAS.372.1496S,2010A&A...513A..37H}). 
\\High resolution observations of CCs showing X-ray surface brightness depressions filled by radio emission suggested that the Active Galactic Nuclei (AGN) in the cores of the brightest cluster galaxies (BCGs) are the key to solve the cooling flow problem. Being fueled by the inflow of cold intra-cluster gas, AGNs inflate radio bubbles which excavate X-ray cavities and drive cocoon shocks in the ICM, prompting a deposition of thermal energy onto the cooling gas and establishing a feedback loop (e.g., \citealt{2007ARA&A..45..117M,2012NJPh...14e5023M,2012AdAst2012E...6G}). 
\\ While such an interplay can explain the achievement of a delicate balance between cooling and heating, the recent observations of dynamical disturbances in CCs raise the issue of whether and how this equilibrium is preserved. Bulk motions of the ICM might result in the displacement of cold gas from the center, thus influencing the cooling cycle \citep{2007PhR...443....1M,2010A&A...516A..32G}.
In fact, the high angular resolution of the \textit{Chandra} space telescope has led to the discovery in a large number of clusters of sharp edges in surface brightness, named cold fronts (e.g., \citealt{2000ApJ...541..542M,2001ApJ...562L.153M}). These contact discontinuities are characterized by a jump in temperature, density and entropy (with the inner side being colder and denser than the outside) and near pressure equilibrium at the interface (see \citealt{2007PhR...443....1M} for a review). 
\\ The origin of cold fronts in relaxed CCs has been identified in minor merger events or off-center passages of small sub-clusters that offset the ICM from the hydrostatic equilibrium \citep{2006ApJ...650..102A}: the perturbation is followed by an oscillating motion (or \textit{sloshing}) which generates one or more discontinuities wrapped around the core in a spiraling geometry. \citet{2010A&A...516A..32G} estimated that cold fronts may exist in the cores of $\sim$2/3 of relaxed clusters; large scale, cold gas spirals associated with cold fronts have actually been observed in CCs (e.g., A2029, \citealt{2013ApJ...773..114P}; Fornax, \citealt{2017ApJ...851...69S}; A2204, \citealt{2017ApJ...838...38C}). Among other models, the turbulent motions of the ICM associated to the sloshing CC have been proposed to power diffuse, non-thermal radio emission associated to the sloshing CC (the so called radio mini-halos, see e.g., \citealt{2013ApJ...762...69Z,2019ApJ...880...70G}).
\\ The presence of dynamical perturbations in CCs could affect the stability of the cooling cycle: first of all, the mechanical energy of the oscillating gas might represent a viable source of heating, provided that it is converted into thermal energy \citep{2003ApJ...590..225C}. Second of all, sloshing displaces the cooled ICM from the center, thus reducing the cooling efficiency and possibly interfering with cold gas deposition onto the BCG (as proposed e.g., in A2495, \citealt{2019ApJ...885..111P}). The combined study of the X-ray emitting ICM and the multi-wavelength conditions of the central AGN in sloshing clusters is essential to provide further information on these topics.
\\ Typically, BCGs host radio-loud AGNs (\citealt{2007MNRAS.379..894B,2009A&A...501..835M,2009ApJ...704.1586S,2015A&A...581A..23K,2015MNRAS.453.1201H}) in the form of Fanaroff-Riley class I radio galaxies (FRI, \citealt{1974MNRAS.167P..31F}), which display an extended morphology with radio lobes reaching distances of several kpc from the center. Extended radio galaxies have been studied in great detail with high-flux limited radio surveys (e.g., the 3C, \citealt{1959MmRAS..68...37E}). \\ Recently, the advent of wide-field surveys in the optical and radio bands (SDSS \footnote{Sloan Digital Sky Survey \citep{2000AJ....120.1579Y}.}, FIRST\footnote{Faint Images of the Radio Sky at Twenty centimeters survey \citep{1995ApJ...450..559B}.}, NVSS\footnote{National Radio Astronomy Observatory Very Large Array (VLA) Sky Survey \citep{1998AJ....115.1693C}.}) allowed to investigate the population of radio galaxies in the mJy regime (e.g., \citealt{2012MNRAS.421.1569B}). These observing programs found that the radio galaxy population in the Local Universe ($z<$ 0.05) is dominated by low-luminosity compact objects characterized by a paucity of extended radio emission (unresolved at the 5'' FIRST resolution). \\ In order to identify sources belonging to this class, \citet{2015A&A...576A..38B} defined as Fanaroff-Riley Type 0 (FR0) any radio galaxy associated with a red massive early type galaxy, with a high mass black hole ($\ge 10^{8} $M$_{\odot}$), spectroscopically classified as low excitation radio galaxy\footnote{A source is classified as low-excitation radio galaxy if the [OIII]$\lambda_{5007}$ equivalent width is $<$10$\AA$ and/or [OII]/[OIII]$>$1 \citep{1997MNRAS.286..241J}.}, and with a radio size of $\le 1-3$ kpc. 
\\High-resolution radio observations of these AGNs have confirmed their compact morphology (e.g., \citealt{2015A&A...576A..38B,2018ApJ...863..155C}, which seems to be maintained also at lower frequencies \citep{2019A&A...631A.176C}. \citet{2018MNRAS.476.5535T} conduced a systematic study in the 2-10 keV band of 19 FR0 galaxies, finding that the accretion rate and X-ray emission of their central engine are similar to those of extended FRIs, excluding that the different radio morphologies can be ascribed to accretion-related differences. 
\\To explain the lack of large scale radio emission, \citet{2015A&A...576A..38B} proposed that the external medium of FR0s could possess peculiar properties (density, clumpiness) capable of decelerating the relativistic jets and preventing its propagation beyond $1-3$ kpc from the center. However, the optical host magnitudes of FR0s are similar to those of FRIs, suggesting that the galaxy-scale ISM environment might not be the answer. From an X-ray point of view, it has been suggested that at least for a fraction of FR0s residing in CCs the cool, dense central ICM could represent another source of frustration for the jets of these radio galaxies. About 50\% of the FR0s in the sample of \citet{2018MNRAS.476.5535T} resides in a dense environment (either a galaxy cluster or a galaxy group), but a direct assessment of the ICM conditions nearby FR0s has never been undertaken. 
\\On the other hand, the environment properties could concur to the radio galaxies' sizes, but not act as a direct cause: if the jet possesses an inner, weak, and short ($< 1$ kpc) relativistic spine, wrapped in a mildly relativistic ($v\sim0.3c$) layer, a dense surrounding medium could be able to quickly decelerate it \citep{2015A&A...576A..38B,2013MNRAS.434.3030B}. The low Lorentz factor of the jet might in turn be associated with the parameters of the central black hole: when the spin and mass assume extreme values, a radio loud AGN produces relativistic and stable jets, capable of excavating their way through the external medium and eventually inflate giant radio lobes \citep{2019MNRAS.482.2294B}. On the contrary, FR0s would be associated with less extreme values of the black hole parameters (relatively lower spin and mass). \\ A final answer on the nature of FR0s has not yet been found: besides investigating the intrinsic jet properties, it is unclear whether the FR0s which reside in galaxy clusters are affected by the condition of the surrounding ICM, and if they are able to establish a feedback loop cycle. 
\\ In this work we focus on the \textit{Chandra} observation of the poorly studied galaxy cluster Abell 795 (A795), at a redshift of $z=0.1374$ \citep{2013ApJ...767...15R}. Previous X-ray observation of this cluster had been performed by the \textit{Rosat All Sky Survey} \citep{1996MNRAS.281..799E}, which measured a flux in the $0.1 - 2.4$ keV band of $F_{\text{X}}=7.1\times10^{-12}$ erg cm$^{-2}$ s$^{-1}$. \citet{2015A&A...575A..48S} investigated the spectroscopic redshifts of the cluster's galaxies, and classified this system as morphologically disturbed; A795 is included in the clusters samples of \citet{2015MNRAS.449..199M} and \citet{2016ApJ...823..116Z}, who confirmed the globally unrelaxed state of the ICM. However, no specific information on this cluster cooling properties and ICM detailed conditions are available. The elliptical BCG J092405.30+14, located at RA, DEC = 09:24:05.3, +14:10:21.5 (J2000) hosts a radio loud AGN. \citet{2018MNRAS.476.5535T} classified it as a FR0 radio galaxy, and concluded that the X-ray incoming photons are produced in the jet and that an inefficient, advection-dominated accretion flow powers the central engine. 
\\Therefore, this cluster offers the opportunity to investigate the possible link between the compact radio morphology of a FR0 and the properties of its hot environment. By performing a detailed analysis of an archival \textit{Chandra} observation of A795 and the BCG J092405.30+14, we probe the thermodynamical condition and cooling efficiency of the ICM. Furthermore, we design a scenario which might confirm or reject the hypothesis that the environment of FR0 radio galaxies affects the jet stability, leading to its inability to move through the external medium, and eventually to its premature disruption. \\ This paper is organized as follows: Section \ref{bcgproperties} describes the optical and radio properties of the BCG, providing a description of the accretion parameters of the AGN; the \textit{Chandra} data reduction is reported in Section \ref{reduction}, while the analysis of the ICM and of the FR0 is detailed in Section \ref{results}. In Section \ref{discussion} we discuss the implications of our results, dedicating our attention to the reasons behind the AGN radio compactness, the possible heating sources in A795, and the outcomes of the ICM complex dynamics. At last, we summarize our conclusions in Section \ref{conclusion}. \\
\linebreak
In this work we adopt the following cosmology: $H_{0}$ = 70 km s$^{-1}$ Mpc$^{-1}$, $\Omega_{\text{m}}$ = 0.3, $\Omega_{\Lambda}$ = 0.7, which results in a conversion between linear and angular scales of 2.43 kpc/'' at A795's redshift. We report every uncertainty at the $1\sigma$ confidence level. The radio spectral index $\alpha$ is defined as $S_{\nu}\propto\nu^{-\alpha}$ (where $\nu$ is the frequency and $S_{\nu}$ is the flux density at the frequency $\nu$).
\section{Radio and optical properties of the central AGN}
\label{bcgproperties}
\begin{figure}
	\centering
	\includegraphics[width=\hsize]{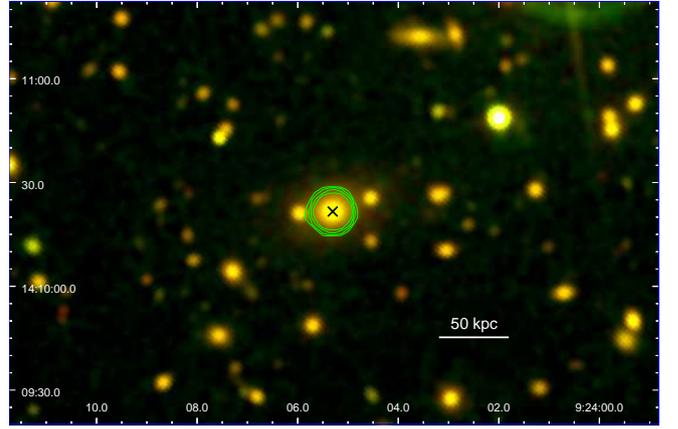}
	\caption{SDSS image of the innermost regions of A795, with 1.4 GHz FIRST radio contours (green) at 5, 10, 15, 20 $\times \sigma$ (with $\sigma$=0.15 mJy). The angular resolution of the FIRST survey is 5''. The black cross indicates J092405.30+14.}
	\label{fig:bcgfirst}
\end{figure}

\citet{2012MNRAS.421.1569B} derived the information on the stellar mass of the BCG (4.2$\times10^{11}$M$_{\odot}$), the 4000\AA-break (1.76) and the [OIII] line luminosity of the AGN ($\sim$5.4$\times10^{40}$ erg s$^{-1}$), arguing that J092405.30+14 is a non-star forming, low-excitation line radio galaxy (LERG). \\ We exploited the values reported by \citet{2012MNRAS.421.1569B} and the velocity dispersion measurement of the SDSS DR12 \citep{2015ApJS..219...12A} to compute the accretion parameters of the central engine in J092405.30+14 (see Tab. \ref{tab:laccledd}): we used the definition of \citet{2004ApJ...613..109H} of $L_{\text{acc}}$=3500$\times L_{\text{[OIII]}}$ to obtain the accretion luminosity (second column); from the M$_{\text{BH}}$-$\sigma_{\star}$ relation of \citet{2002ApJ...574..740T} we derived the mass of the black hole (third column), which we used to compute the Eddington luminosity $L_{\text{Edd}}$ (fourth column). We find a ratio  $L_{\text{acc}}$/$L_{\text{Edd}}$ of $3.8 \times 10^{-3}$, typical of LERG sources in which the central engine is powered by an advection-dominated accretion flow (e.g., \citealt{2018MNRAS.476.5535T}). 
\begin{table}
\centering
	\caption{Parameters of the central engine in J092405.30+14. The $L_{\text{[OIII]}}$ in the first column has been derived by \citet{2012MNRAS.421.1569B}; we obtained the accretion luminosity (second column) using the definition of \citet{2004ApJ...613..109H}:  $L_{\text{acc}}$=3500$L_{\text{[OIII]}}$; we derived the black hole mass in the third column from the stellar velocity dispersion ($\sigma_{\star}= 261\pm9$ km s$^{-1}$, measured by the SDSS DR 12, \citealt{2015ApJS..219...12A}) using the $M_{\text{BH}}$-$\sigma_{\star}$ relation of \citet{2002ApJ...574..740T}; we computed the fourth column as $L_{\text{Edd}}$ $\sim$ 1.3$\times10^{38} M_{\text{BH}}$/M$_{\odot}$ erg s$^{-1}$.}
	\label{tab:laccledd}
		\begin{tabular}{cccc}
			\hline
			 $L_{\text{[OIII]}}$& $L_{\text{acc}}$&  $M_{\text{BH}}$& $L_{\text{Edd}}$\\
			$10^{40}$ erg s$^{-1}$&$10^{44}$ erg s$^{-1}$&  $\times10^{8}$ M$_{\odot}$ &$10^{46}$ erg s$^{-1}$\\
			\hline
			\rule{0pt}{2.5ex}$5.37\pm0.05 $&$1.88\pm0.02 $ &$3.9\pm1.7$ & $4.9\pm2.1$\rule[-1.5ex]{0pt}{2.5ex}\\
			\hline
		\end{tabular}
\end{table}	
\\ Radio observations of J092405.30+14 in the GHz band disclosed its compact radio morphology: it is unresolved in the FIRST images, thus implying an upper limit on its size at 1.4 GHz of 5'' ($\sim$12 kpc, see Fig. \ref{fig:bcgfirst}); \citet{2015MNRAS.453.1201H} did not resolve the source at 4.8 GHz with the \textit{Very Large Array} (VLA) in configuration C (resolution of 3.4''$\sim$8.3 kpc), but resolved it at 8.4 GHz with the VLA in configuration A (resolution of 0.2''$\sim$0.48 kpc). This source has also been observed by \citet{2010MNRAS.408.2261K} with MERLIN at 5 GHz (sub-arcsec resolution), which revealed an elongation in the north-east direction with a largest linear size of 0.30'' ($\sim$0.73 kpc), suggesting a core-jet morphology. A VLBA observation of J092405.30+14 did not detect the source \citep{2015MNRAS.453.1201H}.
\section{\textit{Chandra} observation and data reduction}
\label{reduction}
 A795 was observed on 2010 January 13 by the \textit{Chandra} X-ray telescope (ObsID 11734, P.I. Russell) using the Advanced CCD Imaging Spectrometer S (ACIS-S) in VFAINT mode, with a total exposure time of $\sim 30$ ks. Data reprocessing has been performed with \textit{CIAO 4.12} and \textit{CALDB 4.9.0}: with the \texttt{chandra\_repro} script we performed the bad pixel removal and the instrumental error correction, and with the \texttt{deflare} scripts we removed the background flares. The final exposure time is 29.7 ks.\\ The \textit{CIAO} tool \texttt{wavdetect} has been used to identify point sources in the event file, which we compared to optical reference objects (catalogue \texttt{USNO-A2.0}) to verify that the astrometry of the data is accurate. At last, the \texttt{blanksky} background file corresponding to ObsID 11734 was reprojected to match the observation, and scaled by the hard energy (9-12 keV) count rate of the image. 
 \section{Results}
\label{results}
\subsection{X-ray morphology of the ICM}
\label{morphoA795}
\begin{figure}
	\centering
	\subfloat[]{\includegraphics[width=\hsize]{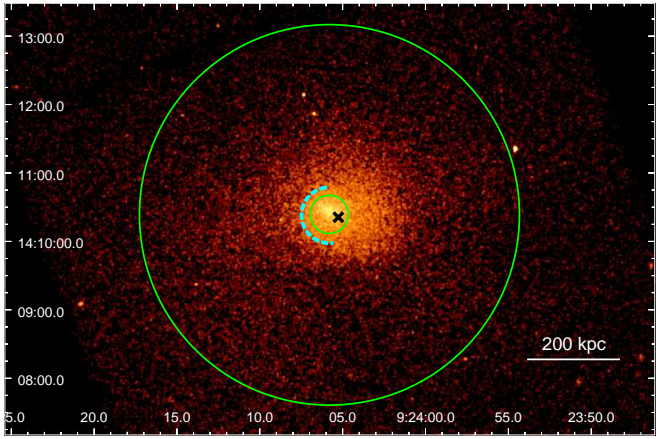}} \\
	\subfloat[]{\includegraphics[width=\hsize]{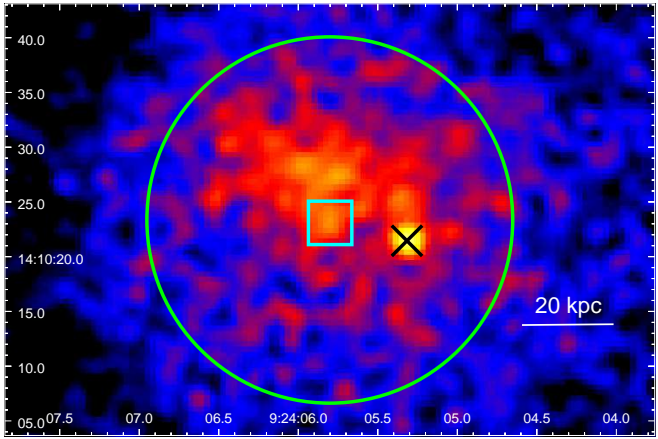}}
	\caption{Panel \textit{a}: background-subtracted, exposure-corrected 0.5 -2 keV image of A795, Gaussian-smoothed with a kernel radius of 1.5''; the outer green circle of radius 167'' ($\sim$400 kpc) has been used to obtain the general properties of the cluster. The cyan arc highlights the best-fit position and extension of the surface brightness discontinuity in the ICM. Panel \textit{b}: 0.5 -2 keV image of the central regions of A795. The image is Gaussian-smoothed with a kernel radius of 1.5''; the cyan box marks the position of the X-ray peak. In both panels, the black cross indicates the central AGN, and the green circle of radius 16.7'' ($\sim$40.6 kpc) corresponds to the innermost bin of the radial spectral analysis (Sect. \ref{spectralA795}).}
	\label{fig:a_chip}
\end{figure}

\begin{figure}
	\centering
	\includegraphics[width=\hsize]{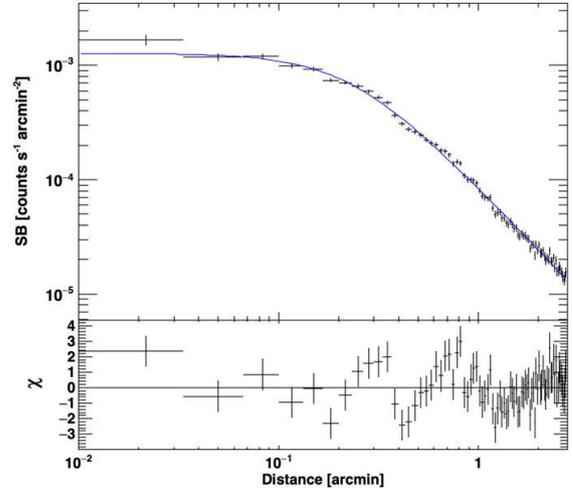}
	\caption{0.5-2 keV surface brightness profile (black points) of A795, fitted with a $\beta$-model (blue line). The lower panel shows the residuals from the best-fit model.}
	\label{fig:profsa}
\end{figure}

\begin{figure}
	\centering
	\subfloat[]{\includegraphics[width=\hsize]{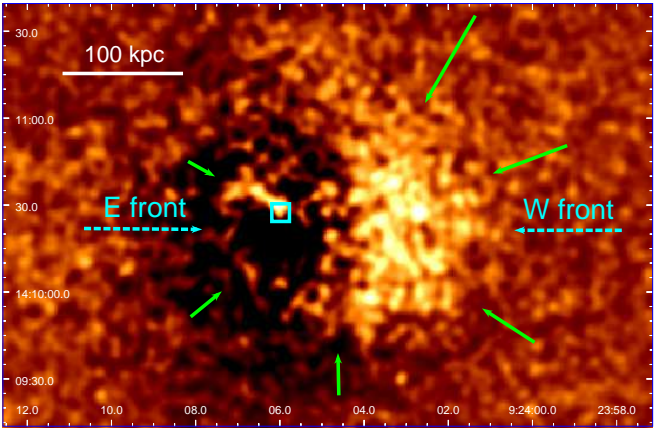}}\\
	\subfloat[]{\includegraphics[width=\hsize]{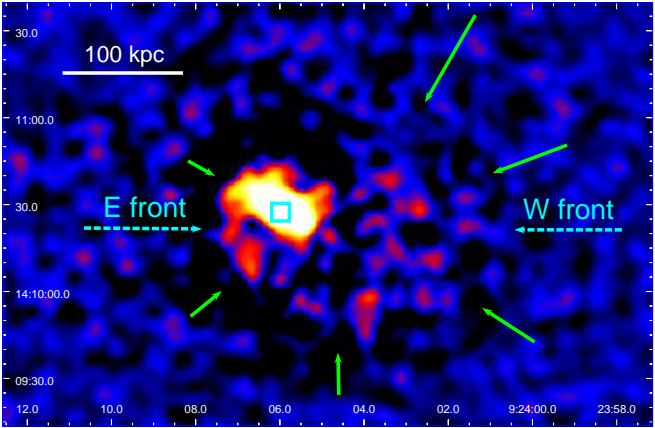}}
	\caption{Panel \textit{a}: 0.5-2 keV residual image of A795, Gaussian-smoothed with a kernel radius of 3 arcsec. Panel \textit{b}: 0.5-2 keV unsharp mask image of A795, obtained by subtracting two images smoothed with a Gaussian of 2.5'' and 10'' axes, respectively. In both panels, the green arrows highlight the spiral geometry, the cyan arrows mark the position of the two surface brightness discontinuities, and the cyan box indicates the X-ray peak.}
	\label{fig:residun}
\end{figure}

\begin{figure}
	\subfloat[\emph{E edge}]{\includegraphics[width=\hsize]{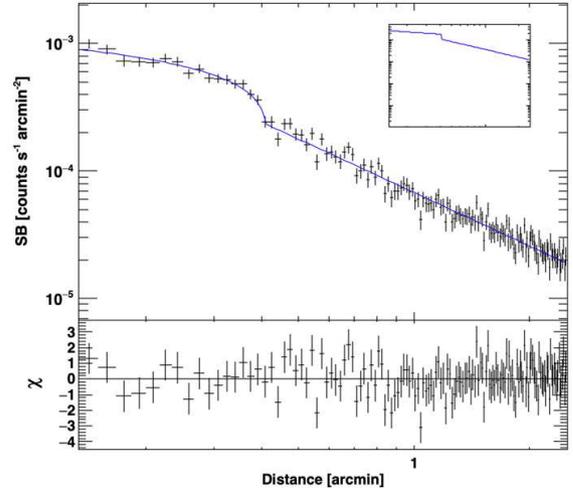}} \\
	\subfloat[\emph{W edge}]{\includegraphics[width=\hsize]{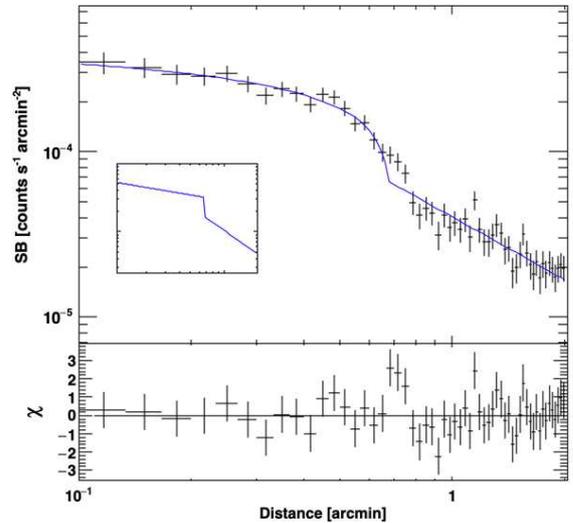}}
	\caption{0.5-2 keV surface brightness profiles along the E (upper panel) and the W (lower panel) discontinuities. In each plot, the best fit broken power-law model is plotted with a blue line over the data and inside a small box.}
	\label{fig:westeast}
\end{figure}

Fig. \ref{fig:a_chip} (panel \textit{a}) shows the background-subtracted, exposure-corrected 0.5-2 keV image of A795, situated in the chip 7 of ACIS-S: we analyzed the morphology of this cluster using the tools of \textit{Proffit-1.5} \citep{2011A&A...526A..79E}, a software designed for the analysis of galaxy cluster X-ray surface brightness profiles.
\\ An inspection of the central regions (Fig. \ref{fig:a_chip}, panel \textit{b}) reveals the presence of an offset of 7.3'' ($\sim$17.7 kpc) between the peak of the X-ray emission (RA, DEC = 09:24:05.8, +14:10:23.3) and the position of the central AGN (RA, DEC = 09:24:05.3, +14:10:21.5).
We excluded the point sources from the surface brightness analysis, so that their emission would not contaminate that of the ICM; the region enclosing the central AGN (derived by using the \texttt{mkpsfmap}, with ecf=0.5, and the \texttt{wavdetect} scripts) is an ellipse of semi-major axis 1.9'' ($\sim$4.6 kpc) and semi-minor axis of 1.7'' ($\sim$4.1 kpc). 
\\ We extracted the surface brightness profile from a series of concentric annuli with bin size of 2'' centered on the X-ray peak and extending to 168'' ($\sim$405 kpc) from the center. The \textit{Proffit} single $\beta$-model \citep{1976A&A....49..137C} provides the best fit to the resulting profile: parameters are reported in Tab. \ref{tab:bestbeta}\footnote{The normalization units (second column of Tab. \ref{tab:bestbeta}) are due to \textit{Proffit} assuming cm$^{2}$=1, even though \textit{Chandra}’s exposure maps have units of cm$^{2}$ s. This inconsistency can be solved by normalizing the exposure map by the ratio between the exposure map value at the aim point and the exposure time, thus loading on \textit{Proffit} an exposure map in units of seconds.}, and the profile is shown in Fig. \ref{fig:profsa}. We found the addition of a second $\beta$-model not to be statistically significant.
\begin{table}
	\caption{Best $\beta$-model fit to the surface brightness profile of A795: (1) parameter $\beta$; (2) profile normalization; (3) core radius; (4) $\chi^{2}$/degrees of freedom.}
	\label{tab:bestbeta}
	\centering
		\begin{tabular}{cccc}
			\hline
			$\beta$& $norm$ & $r_{\text{c}}$ &  $\chi^{2}$/D.o.f\\
			 & cts s$^{-1}$ arcmin$^{-2}$& kpc  & \\
			\hline
			\rule{0pt}{4ex} 0.47$^{+0.01}_{-0.01}$ &  1.28$^{+0.08}_{-0.07}\times10^{-3}$ & 33.5$^{+1.5}_{-1.5}$ & 124.8/80 (1.56) \\
			\hline
		\end{tabular}
\end{table}
\\ \noindent We computed the surface brightness \textit{concentration} $c_{\text{SB}}$ \citep{2008A&A...483...35S} and the \textit{centroid shift} $w$ \citep{1993ApJ...413..492M}, which can be useful diagnostic parameters to obtain information on the cluster's dynamical state. The \textit{concentration} $c_{\text{SB}}$ consists in the ratio between the total fluxes calculated within 40 kpc and 400 kpc from the cluster center:
\begin{equation}
c_{\text{SB}} = \frac{F (r \le 40\,\text{kpc})}{F (r \le 400\,\text{kpc})}
\end{equation}
The \textit{centroid shift w} assesses the shift of the position of the X-ray centroid when changing the used aperture from a radius $R_{\text{max}}$ to the X-ray peak. This parameter is defined as follows: 
\begin{equation}
w = \frac{1}{R_{\text{max}}} \sqrt{\frac{\Sigma(\Delta_{\text{i}} - \langle\Delta\rangle)^{2}}{N-1}}
\end{equation}
\noindent where $N$ is the number of considered apertures, and $\Delta_{\text{i}}$ is the distance between the $i$-th centroid and the one obtained with aperture $R_{\text{max}}$ (set at 400 kpc to be consistent with $c_{\text{SB}}$).
\\We obtained $w=0.03$ and $c_{\text{SB}}=0.15$: these values place A795 at the boundary between strong CC clusters ($w>0.03$ and $c_{\text{SB}}>0.16$) and NCC clusters ($w<0.03$ and $c_{\text{SB}}<0.08$). We note that this result hints at the presence of dynamical disturbances in the ICM, that could be responsible for the observed offset between the X-ray peak and the BCG. Moreover, the prevalence of a single $\beta$-model for the surface brightness profile suggests that A795 might not be a strong CC cluster, which typically need a second $\beta$-model component to fit the surface brightness (e.g., \citealt{1999ApJ...517..627M}); with the spectral analysis of Subsect. \ref{spectralA795}, we further investigate this argument. 
\\ The undulating pattern of the residuals from the $\beta$-model fit (Fig. \ref{fig:profsa}) might be the outcome of azimuthally averaging surface brightness discontinuities in the ICM. Indeed, the 0.5-2 keV image of Fig. \ref{fig:a_chip} (panel \textit{a}) reveals a sharp surface brightness discontinuity on the east side of the cluster. To better highlight this and possibly other substructures, we produced a residual image (shown in Fig. \ref{fig:residun}, panel \textit{a}) with the \texttt{savedeviations} tool of \textit{Proffit}, which computes the pixel-by-pixel residuals (in units of $\sigma$) between the data and the best fit model to the surface brightness profile (Tab. \ref{tab:bestbeta}). This technique unveils the presence of a large scale spiral in the ICM: starting from the cluster's center it bends to south-east following the identified discontinuity, then proceeds to north-west reaching a distance of $\approx$180 kpc from the X-ray peak. We note that the residual image reveals the presence of another discontinuity in the ICM west to the cluster center: this feature is less sharp than the eastern one, but appears more extended. 
\\ We show in Fig. \ref{fig:residun} (panel \textit{b}) the unsharp mask image of A795, obtained by subtracting two images of the cluster (smoothed with a Gaussian of 2.5'' and 10'' axes, respectively); the image shows a spiraling feature, which appears very similar in geometry and extension to the one identified in the residual image.
\\ To obtain the position and density ratio of the two jumps we extracted surface brightness profiles across the east (E) and west (W) discontinuities, using circular sectors of bin size 2'' with opening angles 180$^{\circ}$ for the E jump and 207$^{\circ}$ for the W jump. The surface brightness profiles were fitted using a broken power-law model (\texttt{bknpow}) with a density jump numerically projected along the line of sight.
The free parameters of this model are the slopes before and after the jump ($\alpha_{1}$ and $\alpha_{2}$) the normalization of the profile, the position of the jump (\textit{cutrad}), and the density jump (\textit{jump}). Fig. \ref{fig:westeast} and Tab. \ref{tab:frontsNEW} report the fitted profiles and the best fit parameters, respectively; we highlight the following results:
\begin{enumerate}
	\item Front E is located $\sim$60 kpc from the center, and corresponds to a density jump of $1.69\pm0.07$; the very high significance of the jump ($\approx 10\sigma$) is consistent with a sharp discontinuity.
	\item Front W is located at $\sim$178 kpc from the center, and corresponds to a density jump of $1.93^{+0.31}_{-0.26}$. This front has a lower significance ($\approx 3\sigma$): the residuals of the profile in Fig. \ref{fig:westeast} (panel \textit{b}) indeed show that the discontinuity is not as sharp as the east one. 
\end{enumerate}
\begin{table*}
\centering
	\caption{Properties and best fit parameters of the \texttt{bknpow} model used to describe the surface brightness profile of the E and W edges: (1) name of the edge; (2) distance of each discontinuity from the X-ray peak; (3) slope of the first power-law; (4) slope of the second power-law; (5) position of the jump referred to the center of the sectors; (6) profile normalization; (7) density jump factor; (8) $\chi^{2}/$degrees of freedom.}
	\label{tab:frontsNEW}
	\begin{tabular}{ c c c c c c c c }
		\hline
		 Edge & D  & $\alpha_{1}$ & $\alpha_{2}$ & $cutrad$ & $norm$& $jump$& $\chi^{2}/$D.o.f. \\
		  &  kpc ('') &  &  & arcmin & $10^{-3}$ cts s$^{-1}$ arcmin$^{-2}$ & (${n_{\text{e}}}$)&  \\
		\hline
		\rule{0pt}{4ex}  E & 59.6 (24.5)&   0.35$\pm 0.10$ & 1.20$\pm 0.01$  & 0.409 $\pm 0.001$& 1.33$\pm 0.09$ &  1.69$\pm 0.07$ & 139.1/137 (1.02)\\
		\rule{0pt}{4ex}  W & 178.2 (73.4) &  0.28$\pm 0.09$ & 1.13$\pm 0.05$  & 0.667 $\pm 0.015$& 0.28$\pm 0.03$ &  1.93$\pm 0.31$& 53.3/52 (1.03)\\
		\hline
	\end{tabular}
\end{table*}
\indent We argue that the large-scale geometry of the ICM might be the outcome of a \textit{sloshing} mechanism, comparable to that observed in other clusters which display spiral morphology (e.g., A2142 \citealt{2000ApJ...541..542M}; A1795 \citealt{2001ApJ...562L.153M}; A2029, \citealt{2013ApJ...773..114P}), which could explain the offset between the X-ray peak and the BCG. Considering an original \textit{relaxed} configuration, with a pronounced central density peak typical of CC clusters, the displacement of the ICM could have stretched the density peak and explain the single $\beta$-model prevalence for the surface brightness profile. Moreover, the two surface brightness jumps nicely follow the spiral feature, suggesting that the two discontinuities could be cold fronts. Testing this hypothesis requires a spectroscopic confirmation: we produced temperature, density and pressure profiles along the two edges (Subsect. \ref{specdis}).
\subsection{Spectral analysis of the ICM}
\label{spectralA795}
\begin{figure}
	\centering
	\subfloat[]{\includegraphics[width=\hsize]{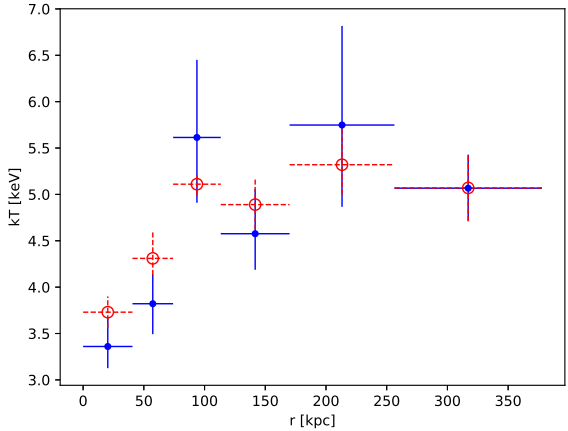}} \\
	\subfloat[]{\includegraphics[width=\hsize]{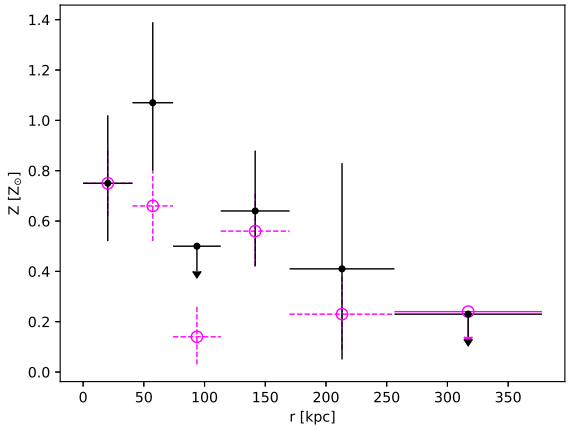}}
	\caption{Panel \textit{a}: Projected (red, dashed) and deprojected (blue, solid) temperature profile for A795. Panel \textit{b}: Projected (pink, dashed) and deprojected (black, solid) abundance profile for A795. Both profiles are centered on the X-ray peak; errorbars on the x-axis represent the bin width.}
	\label{fig:deprojtempab}
\end{figure}

\begin{figure*}
	\centering
	\subfloat[Density]{\includegraphics[width=0.33\hsize]{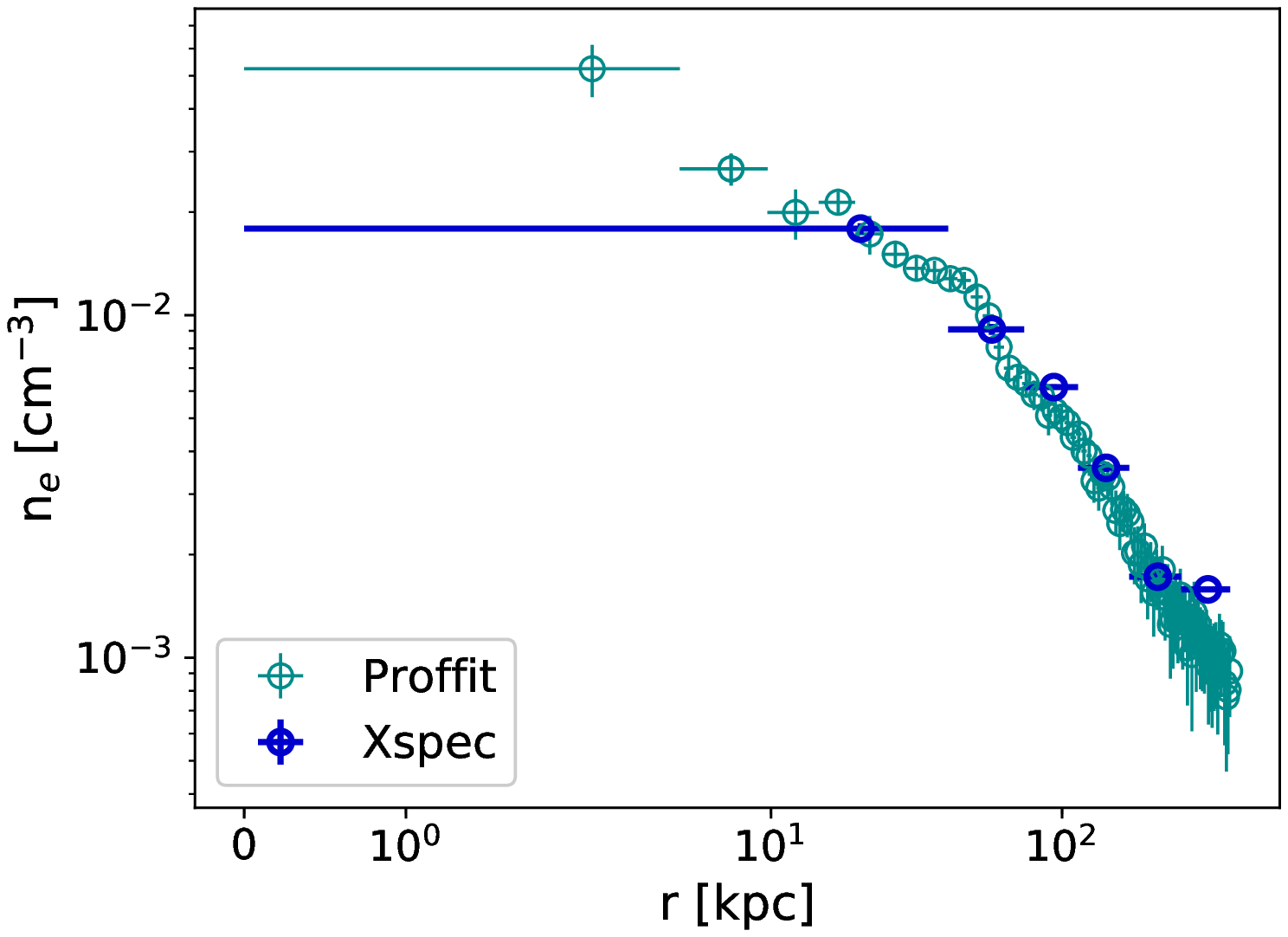}} 
	\subfloat[Pressure]{\includegraphics[width=0.33\hsize]{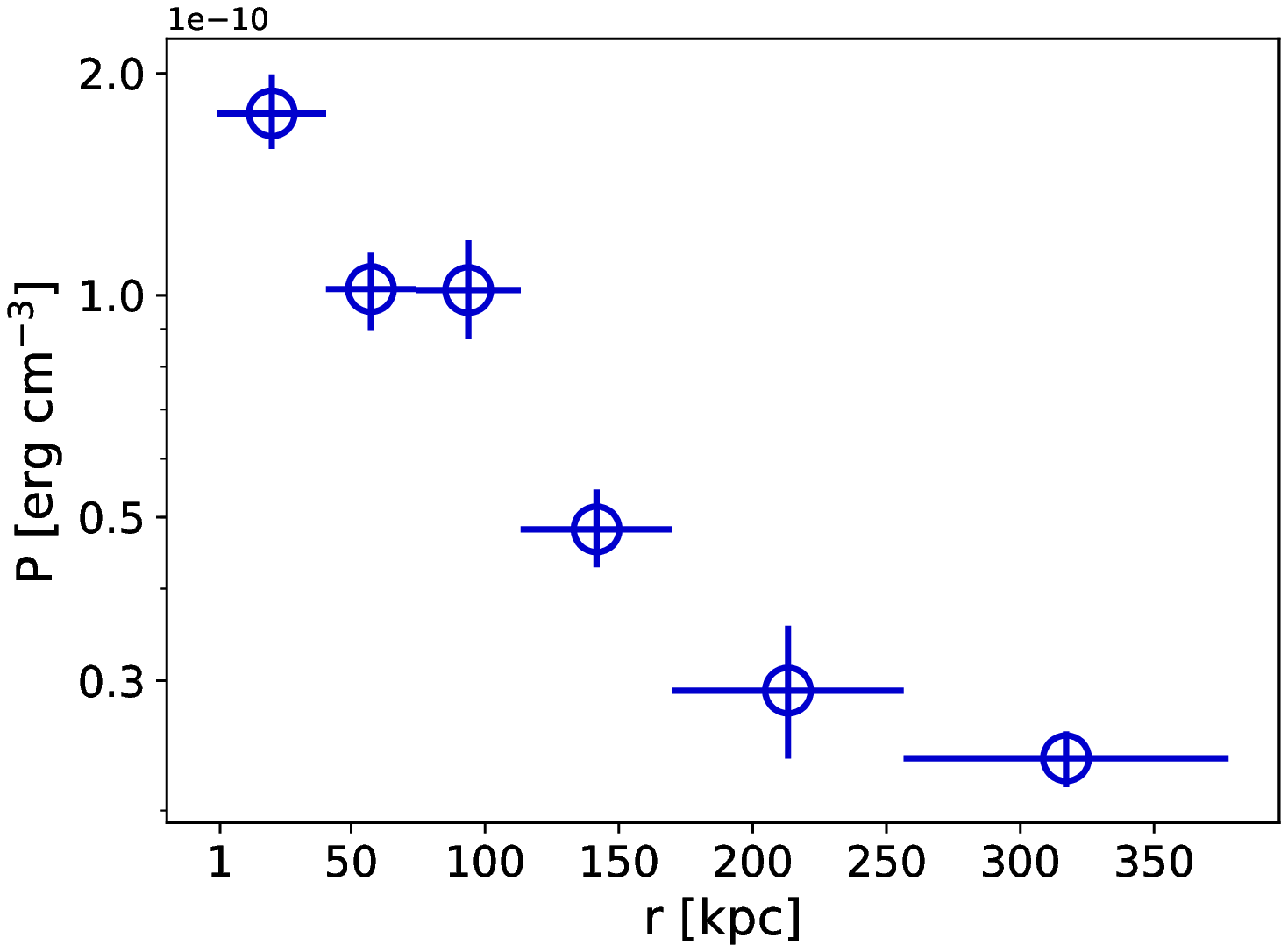}} 
	\subfloat[Entropy]{\includegraphics[width=0.33\hsize]{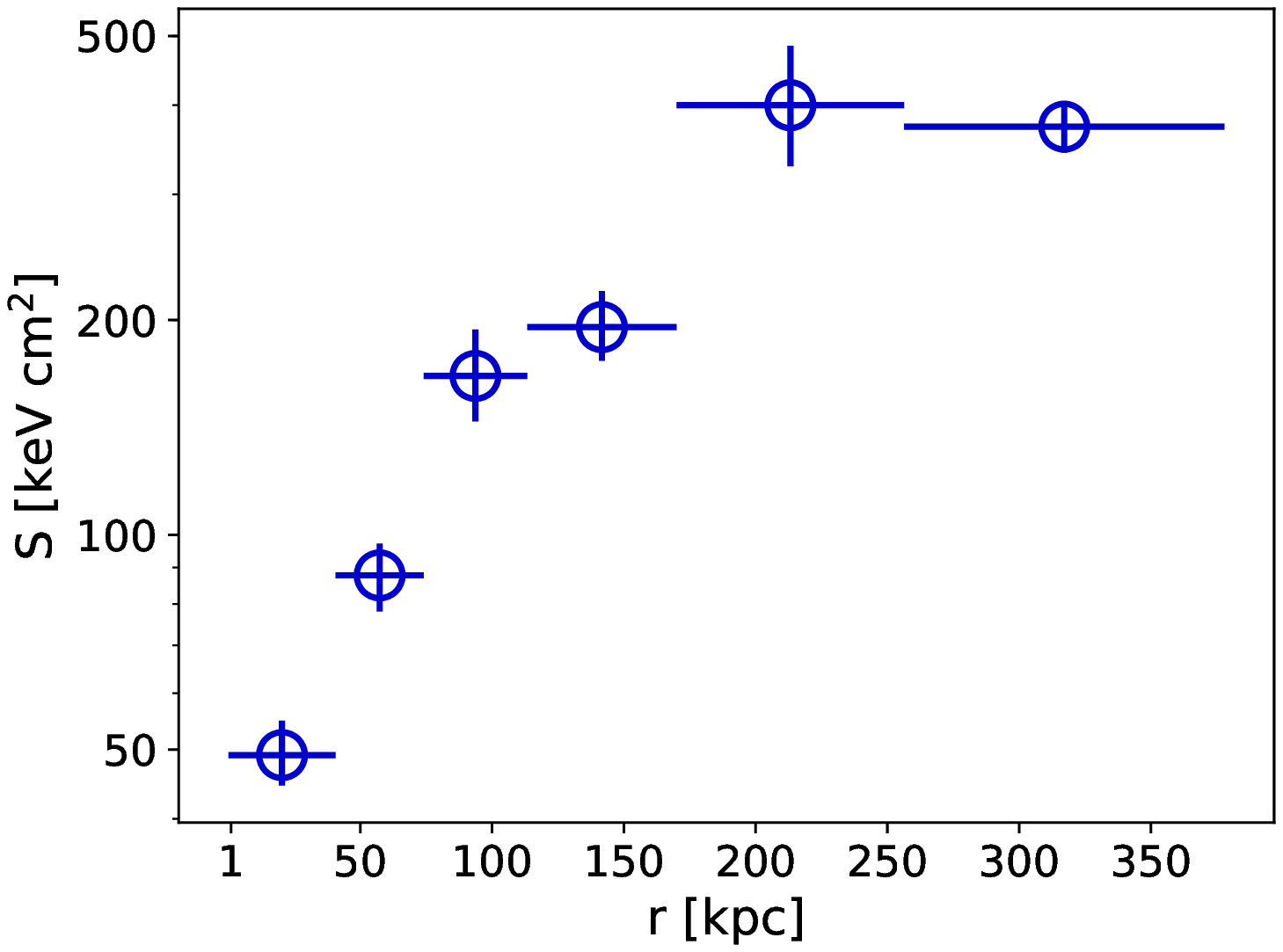}} 
	\caption{Radial profiles of density (panel \textit{a}), pressure (panel \textit{b}) and entropy (panel \textit{c}) for A795 (blue points), obtained from the deprojected spectral analysis. The turquoise points in panel \textit{a} represent the density profile obtained by deprojecting the surface brightness profile.}
	\label{fig:densentpres}
\end{figure*}

In order to produce a detailed characterization of the thermodynamical properties of A795, and to verify the classification of this object as a weakly CC system, we performed a spectral analysis of the ICM. Spectral fitting was performed in the energy range 0.5-7 keV with \textit{Xspec - v.12.10}; the background spectrum has been extracted from the blanksky event file in the region of interest, and has been subtracted before the fitting procedure. For every thermal model and photoelectric absorption model employed in this work, we used the table of abundances of \citet{2009ARA&A..47..481A}.
\\ The global spectral properties of A795 were deduced from a circular region centered on the X-ray peak and covering the entire chip 3: the resulting circle (shown in Fig. \ref{fig:a_chip}) has a radius of 167'' ($\sim$405 kpc); as for the surface brightness analysis, point sources were excluded. We fitted the spectrum with a \texttt{tbabs$\ast$apec} model: the galactic absorption was fixed at the value $N_{\text{H}} \approx 2.89 \times 10^{20}$ cm$^{-2}$ \citep{2016A&A...594A.116H}; the second term accounts for the emission of a collisional ionized gas. The redshift ($z\sim$0.137) was fixed, while $kT$ (temperature), $Z$ (metal abundance in units of Z$_{\odot}$) and $norm$ (normalization of the spectrum) were left free to vary. We measured $kT=$4.63$\pm 0.12$ keV, $Z=$0.38$\pm0.05$ Z$_{\odot}$, $F(0.5-7\,\text{keV}) = 6.95^{+0.04}_{-0.04}\times10^{-12}$ erg s$^{-1}$ cm$^{-2}$, and $L(0.5-7\,\text{keV}) = 3.43^{+0.04}_{-0.03}\times10^{44}$ erg s$^{-1}$ (the $\chi^{2}$/D.o.f. is 375/325). 
\\ To investigate the properties of the ICM at different distances from the center, we produced radial profiles of thermodynamic variables by extracting a spectrum from six concentric regions which contained at least 4000 net counts centered on the X-ray peak and extending to $\approx$380 kpc from the cluster center. We performed a projected analysis by fitting each spectrum with a \texttt{tbabs$\ast$apec} model, with $N_{\text{H}}$ fixed at the value of $2.89\times10^{20}$ cm$^{-2}$, and redshift fixed at 0.1374. Tab. \ref{tab:radialproj} lists the best-fit results, and Fig. \ref{fig:deprojtempab} shows the projected temperature profile (red dashed, panel \textit{a}), and the projected metallicity profile (pink dashed, panel \textit{b}). 
\begin{table}
\centering
	\caption{Fit results of the projected radial analysis of A795: (1) outer radius of each annulus (the inner radius is the outer radius of the previous annulus); (2) net photon counts (fraction w.r.t. the total counts from the same region); (3) temperature; (4) metallicity; (5) $\chi^{2}$/degrees of freedom.}
		\label{tab:radialproj}
		\begin{tabular}{lcccr}
			\hline
			R$_{\text{o}}$& Counts & $kT$ & $Z$ &  $\chi^{2}$/D.o.f. \\
			kpc &  &  keV & Z$_{\odot}$ &  \\
			\hline
			40.6 & 4298 (98.8\%) & 3.73$^{+0.17}_{-0.17}$  & 0.75$^{+0.14}_{-0.13}$ & 122.9/114  \\
			74.1& 4542 (97.6\%) & 4.31$^{+0.28}_{-0.18}$  & 0.66$^{+0.14}_{-0.14}$ & 105.2/127 \\
			113.3 & 4497 (95.6\%) & 5.11$^{+0.12}_{-0.12}$  & 0.14$^{+0.12}_{-0.11}$ & 135.5/131 \\
			170.0 & 4811 (91.2\%) & 4.89$^{+0.27}_{-0.27}$  & 0.56$^{+0.15}_{-0.14}$ & 136.9/139 \\
			256.4 & 4780 (82.1\%) & 5.32$^{+0.39}_{-0.32}$  & 0.23$^{+0.15}_{-0.14}$ & 156.1/156 \\
			377.8 & 5337 (70.9\%) & 5.07$^{+0.36}_{-0.36}$  & $<$ 0.24                        & 178.2/189 \\
			\hline
		\end{tabular}
\end{table}	
Thawing the column density parameter $N_{\text{H}}$ did not provide significant variations in the results: either the column density reached a value consistent within errors with the fixed value, or the $\chi^{2}$/D.o.f. did not indicate a significant improvement. \\ To account for the possible presence of multi-phase gas, we considered the possibility of adding a second thermal component, by fitting each annulus with a \texttt{tbabs$\ast$(apec$+$apec)} model: again, we did not find a significant improvement in any annulus, as either the normalization of the second thermal component was negligible with respect to that of the first one, or the $\chi^{2}$/D.o.f. improvement was not statistically significant.
\\ In order to remove the contribution of the ICM along the line of sight, we fitted the same spectra with a \texttt{projct$\ast$tbabs$\ast$apec} model, where the first component performs a 3-D to 2-D projection of ellipsoidal shells onto elliptical annuli. The best fit results are listed in Tab. \ref{tab:radialdeproj} ($\chi^{2}$/D.o.f. is 832.7/856); Fig. \ref{fig:deprojtempab} shows the temperature (blue solid, upper panel) and metallicity (black solid, lower panel) profiles.
\\ The temperature drop in the inner annuli suggests a higher cooling efficiency in the central regions, and the decreasing metallicity profile is consistent with a heavier central enrichment, as typically observed in relaxed clusters (see e.g., \citealt{2019MNRAS.483..540L}). We also note that the deprojected metallicity profile (black points in Fig. \ref{fig:deprojtempab}, panel \textit{b}) hints at higher abundances in the second and fourth annuli, where the east and west discontinuity are located, respectively: this could suggest a sloshing-related redistribution of the enriched ICM. However, probing the spatial distribution of metallicity in A795 would require deeper X-ray observations: due to the large errorbars of the deprojected profile no firm conclusion can be drawn with the current exposure.
\\The deprojected analysis is also fundamental in terms of deriving the electron density of the plasma. The \textit{norm} parameter of the \texttt{apec} model is defined as: 
\begin{equation}
norm = \frac{10^{-14}}{4\pi[D_{\text{A}}(1+z)]^{2}} \int n_{\text{e}}n_{\text{p}}dV
\label{normapec}
\end{equation}
\noindent where $D_{\text{A}}$ is the angular distance from the source, $z$ is the redshift, $n_{\text{e}}$ and $n_{\text{p}}$ are the electron and proton densities, and $V$ is the projected volume of the emitting region. By reverting Eq. \ref{normapec} and assuming $n_{\text{e}}\sim1.2n_{\text{p}}$ (e.g., \citealt{2012AdAst2012E...6G}), it is possible to estimate the electron density $n_{\text{e}}$ as:
\begin{equation}
n_{\text{e}} = \sqrt{10^{14}\bigg(\frac{4\pi \times norm \times[D_{\text{A}}(1+z)]^{2}}{0.83 V}\bigg)}
\label{nenorm}
\end{equation}
\noindent where $V$ is the volume of the spherical shells and \textit{norm} is the normalization of the deprojected \texttt{apec} component (when the \texttt{projct} model is used). \\We computed the electron density of each annulus: results are presented in Tab. \ref{tab:radialdeproj}, where we also report pressure ($P=1.83n_{\text{e}}kT$) and entropy ($K=kT/n_{\text{e}}^{2/3}$) of the ICM. The density, pressure, and entropy profiles of A795 obtained with this method are shown in Fig. \ref{fig:densentpres} (blue points). 
\begin{table*}
	\centering
	\caption{Deprojected spectral analysis: (1) outer radius of each annulus (the inner radius is the outer radius of the previous annulus); (2) net photon counts (fraction w.r.t. the total counts from the same region); (3) temperature; (4) metallicity; (5) profile normalization; (6) electron density; (7) pressure of the ICM; (8) entropy of the ICM.}
	\label{tab:radialdeproj}
		\begin{tabular}{cccccccc}
			\hline
			R$_{\text{o}}$& Counts&$kT$ & $Z$  & $norm$   & $n_{\text{e}}$ & $P_{\text{ICM}}$& $K_{\text{ICM}}$\\
			kpc && keV & Z$_{\odot}$ & $10^{-4}$  & 10$^{-3}$cm$^{-3}$ & 10$^{-10}$erg cm$^{-3}$ & keV cm$^{2}$\\
			\hline
			\rule{0pt}{4ex} 40.6 & 4298 (98.8\%) & 3.36$^{+0.31}_{-0.24}$  & 0.75$^{+0.27}_{-0.23}$ & 5.57$^{+0.40}_{-0.40}$ & 17.89$^{+0.06}_{-0.06}$ & 1.77$^{+0.23}_{-0.19}$ & 49.12 $\pm 5.78$\\
			74.1& 4542 (97.6\%) & 3.82$^{+0.32}_{-0.33}$  & 1.07$^{+0.32}_{-0.27}$ & 7.29$^{+0.53}_{-0.53}$ & 9.09$^{+0.33}_{-0.33}$ & 1.02$^{+0.12}_{-0.12}$ & 87.75 $\pm 9.47$\\
			113.3 & 4497 (95.6\%) & 5.61$^{+0.84}_{-0.70}$  & $<$ 0.5 & 10.36$^{+0.39}_{-0.35}$ & 6.16$^{+0.12}_{-0.10}$ & 1.02$^{+0.17}_{-0.14}$ & 166.98 $\pm 26.98$\\
			170.0 & 4811 (91.2\%) & 4.58$^{+0.48}_{-0.39}$  & 0.64$^{+0.24}_{-0.22}$ & 11.53$^{+0.63}_{-0.62}$ & 3.58$^{+0.09}_{-0.09}$ & 0.48$^{+0.06}_{-0.05}$ & 195.48 $\pm 24.14$\\
			256.4 & 4780 (82.1\%) & 5.75$^{+1.07}_{-0.88}$  & 0.41$^{+0.42}_{-0.36}$ & 9.21$^{+0.70}_{-0.70}$ & 1.72$^{+0.07}_{-0.07}$ & 0.29$^{+0.07}_{-0.06}$ & 399.94 $\pm 84.55$\\
			377.8 & 5337 (70.9\%) & 5.07$^{+0.36}_{-0.35}$  & $<$ 0.23                        & 24.11$^{+0.78}_{-0.79}$ & 1.58$^{+0.03}_{-0.03}$ & 0.24$^{+0.02}_{-0.02}$ & 373.22 $\pm 30.70$\\
			\hline
		\end{tabular}
\end{table*}	
\\We adopted a different method to obtain the density of the ICM, based on the deprojection of the surface brightness profile. In particular, we used the \textit{Proffit} tools \texttt{deproject} and \texttt{density} to deproject the surface brightness profile of Fig. \ref{fig:profsa}, by providing a conversion factor defined as the ratio between the count rate (0.985 cts/s) and the \textit{norm} parameter of the \texttt{apec} which describes the plasma. As the normalization parameter differs from one annulus to another, we have chosen to use the normalization of the overall spectrum within $\sim$405 kpc (7.25$\times10^{-3}$). Fig. \ref{fig:densentpres} (turquoise points in panel \textit{a}) shows the density profiles obtained with this method, which is consistent with that derived from the spectral analysis, considering the different resolutions (particularly in the central region). 
\subsection{Cooling properties of A795}
\label{coolingprop}
\begin{figure}
	\centering
	\includegraphics[width=\hsize]{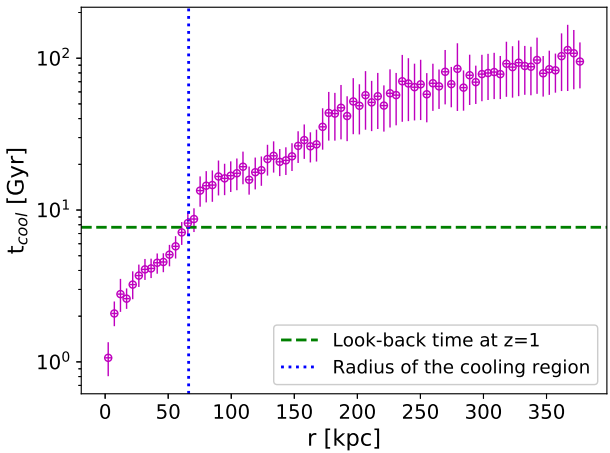} 
	\caption{Cooling time profile of A795. The green, dashed horizontal line corresponds to $t_{\text{cool}} = 7.7$ Gyr; the blue, dotted vertical line at $r=66.2$ kpc indicates the cooling radius. The cooling time profile is consistent with the central $t_{\text{cool}}$ reported in \citet{2016ApJ...823..116Z}.}
	\label{fig:tcoolprofilo}
\end{figure}

In order to determine whether and how cooling is acting in A795, we combined the spectroscopically-determined temperature and the high resolution density profiles to compute the cooling time of the ICM as:
\begin{equation}
t_{\text{cool}} = \frac{\gamma}{\gamma -1} \frac{kT}{\mu X n_{\text{e}}\Lambda(T)}
\label{tcooleq}
\end{equation}
\noindent where $n_{\text{e}}$ is the electron number density, $\gamma = 5/3$ is the adiabatic index, $\mu \approxeq 0.6$ is the molecular weight, $X \approx 0.7 $ is the hydrogen mass fraction and $\Lambda(T)$ is the cooling function \citep{1993ApJS...88..253S}. The cooling time profile for A795 is presented in Fig. \ref{fig:tcoolprofilo}. 
\\ It is possible to define the cooling radius $r_{\text{cool}}$ as the radius at which $t_{\text{cool}}$ is less than the time for which the system has been relaxed, usually assumed to be the look-back time at $z=1$ (that is approximately $7.7 $ Gyr) (e.g., \citealt{2004ApJ...607..800B,2012AdAst2012E...6G}). For A795 we measure a cooling radius of $r_{\text{cool}} = 27.3 \pm 1.3\,\text{arcsec} = 66.2 \pm 3.2\,\text{kpc}$\footnote{To obtain $r_{\text{cool}}$ we fitted the cooling time profile with a power-law relation using the \textit{Bivariate Correlated Errors and intrinsic Scatter} (BCES, \citealt{1996ApJ...470..706A}) library of \textit{Python3.5.4} with the $y/x$ method, and selected $r_{\text{cool}}$ as the intersection between the best fit power-law and $t_{\text{cool}} = 7.7$ Gyr.}. We note that the central cooling time  $1 < t_{\text{cool}}[\text{Gyr}] < 7.7$ provides a further evidence to classify A795 as a (weekly) CC cluster. We also report that \citet{2010A&A...513A..37H} claimed that strong CC clusters display $t_{\text{cool}}[\text{Gyr}] < 1$; this does not hold for A795, but our central radial bin has a cooling time of $1.06_{-0.26}^{+0.28}$ Gyr and its surface brightness is not consistent with the global single $\beta$-model profile of Fig. \ref{fig:profsa}: within errors only the X-ray peak might be at the boundary between strong and weak CCs.
\linebreak
\\ \noindent We performed a spectral analysis of the emission inside the cooling region (defined as a circular region centered on the X-ray peak and with radius equal to $r_{\text{cool}}$) to estimate the cooling luminosity and the mass deposition rate $\dot{M}$. To fulfill this purpose, we followed four approaches:
\begin{enumerate}
    \item We estimated the deprojected luminosity of the cooling region by extracting the spectrum of two concentric annuli, the first being the cooling region and the second extending from $r_{\text{cool}}$ to the edge of the chip 3. Assuming that the emission from the cooling region can be described by a single thermal model, we fitted the spectra with a \texttt{projct$\ast$tbabs$\ast$apec} model and computed the 0.1-100 keV luminosity of the \texttt{apec} component within $r_{\text{cool}}$.
	\item Secondly, we fitted the spectrum of the cooling region with a \texttt{tbabs$\ast$(apec+mkcflow)} model: as an approximation, in this method the \texttt{mkcflow} and \texttt{apec} components are intended to describe the gas within the cooling region and the ambient cluster gas along the line of sight, respectively. The parameters of the thermal component were left free to vary; the abundance and $kT_{\text{high}}$ of the \texttt{mkcflow} were tied to those of the \texttt{apec} component, while $kT_{\text{low}}$ was fixed to the minimum allowed value (0.0808 keV). The normalization of the \texttt{mkcflow} (corresponding to the \textit{mass deposition rate}) was left free to vary. We computed the bolometric luminosities in the 0.1-100 keV band of the \texttt{mkcflow} and the \texttt{apec} components separately (using the \texttt{editmod} command on \textit{Xspec}).
	\item To combine the above two methods we fitted the spectra of the two concentric annuli (used in the first method) with a \texttt{projct$\ast$tbabs$\ast$(apec$+$mkcflow)} model. We left free the abundance, the temperature and the normalization of the \texttt{apec} component in each region. The abundance and $kT_{\text{high}}$ of the \texttt{mkcflow} were tied to the outer thermal model, while $kT_{\text{low}}$ was fixed to the minimum allowed value (0.0808 keV). The normalization of the \texttt{mkcflow} was allowed to vary inside $r_{\text{cool}}$, while it was fixed to zero outside $r_{\text{cool}}$, where no efficient cooling is expected.
	\item At last, we tried a numerical approach: the bolometric X-ray luminosity can be defined as:
	\begin{equation}
	L_{\text{X}} = \int_{V} n_{\text{e}} n_{\text{i}}\Lambda(T) dV,
	\end{equation}
	where $n_{\text{i}}$ is the ions density, $\Lambda(T)$ is the cooling function and $V$ is the volume over which the luminosity is calculated; considering spherical shells of width $dr$, it is possible to write $dV = 4\pi r^{2} dr$. By making direct use of the \textit{Proffit} density profile (Fig. \ref{fig:densentpres}, panel \textit{a}), we performed the integration over the same spherical shells used for the surface brightness profile (i.e. with $dr$ = 2'' = 4.86 kpc) inside the cooling region. 
\end{enumerate}
\begin{table*}
	\centering
	\caption{Results of the spectral analysis of the cooling region of A795: (1) four methods used to study the emission of the cooling region, namely: a deprojected \texttt{apec} component (Method i), a \texttt{mkcflow} component plus an \texttt{apec} one to describe the ambient gas (Method ii), a deprojection of a combined \texttt{apec$+$mkcflow} model (Method iii), and a numerical integration of the density profile (Method iv); (2) inner and outer radius of the cooling region and (when a deprojection was performed) of the outer annulus; (3) temperature; (4) metallicity; (5) normalization of the \texttt{mkcflow} component used in methods 1 and 3; (6) $\chi^{2}$/degrees of freedom; (7) different estimates of the bolometric luminosity within $r_{\text{cool}}$: for Method ii and Method iii the first entry reports the \texttt{mkcflow} luminosity, the second entry reports the \texttt{apec} luminosity (second entry); (8) classical mass deposition rate $\dot{M}_{\text{CF}}$, computed using Eq. \ref{dotmlcool}: the value for Method 3 corresponds to the sum of the \texttt{apec} and \texttt{mkcflow} luminosities.}
	\label{tab:coolmethod}
		\begin{tabular}{cccccccc}
			\hline
			& R$_{i}$ - R$_{o}$ & $kT$ & $Z$  & $\dot{M}_{\text{obs}}$ & $\chi^{2}$/D.o.f.  & $L_{\text{bol}}$ & $\dot{M}_{\text{CF}}$\\
			&  kpc & keV & Z$_{\odot}$ & M$_{\odot}$ yr$^{-1}$ &  & 10$^{44}$erg s$^{-1}$  & M$_{\odot}$ yr$^{-1}$ \\
			\hline
			\multirow{2}{*}{Method i} & 0-66.2 & 3.87$^{+0.14}_{-0.14}$  & 0.74$^{+0.11}_{-0.10}$ & & 485/465 (1.04) & 1.07 $^{+0.06}_{-0.06}$& 108.7$\pm6.1$ \\
			& 66.2-407.2 & 5.03$^{+0.15}_{-0.15}$  & 0.28$^{+0.06}_{-0.06}$  & & &   &\\
			\hline
		    Method ii& 0-66.2  & 4.12$^{+0.22}_{-0.19}$  & 0.72$^{+0.10}_{-0.09}$  & $7.16^{+6.30}_{-6.26}$ & 178/157 (1.13) & \begin{tabular}{@{}c@{}}  0.05$^{+0.02}_{-0.02}$ \\ 1.16$^{+0.06}_{-0.06}$\end{tabular} 
			& \\
			\hline
			\multirow{2}{*}{Method iii} & 0-66.2 & 4.04$^{+0.23}_{-0.21}$  & 0.77$^{+0.12}_{-0.11}$ & $7.00^{+6.13}_{-6.02}$& 484/464 (1.04) & 
			\begin{tabular}{@{}c@{}}   0.05$^{+0.02}_{-0.02}$ \\ 1.03$^{+0.05}_{-0.05}$ \\  \end{tabular} 
			& 108.8$\pm5.8$ \\
			& 66.2-407.2 & 5.03$^{+0.15}_{-0.15}$  & 0.28$^{+0.06}_{-0.06}$  &  & &   &\\
			\hline
			Method iv & 0-66.2 & & & &   & $0.93^{+0.07}_{-0.07}$ & 91.8$\pm$10.6\\
			\hline
		\end{tabular}
\end{table*}
We proceeded to evaluate the cooling rate expected by these luminosities: the cooling flow classical model predicts that the power emitted within the cooling region $L_{\text{cool}}$ is related to the amount of matter which crosses $r_{\text{cool}}$ through:
\begin{equation}
\dot{M}_{\text{CF}} = \frac{2}{5}\frac{\mu m_{\text{p}}}{kT} L_{\text{cool}}
\label{dotmlcool}
\end{equation}
\noindent We used this equation to compute the $\dot{M}_{\text{CF}}$ predicted by each of the above methods: the estimates are reported in the last column of Tab. \ref{tab:coolmethod}. We highlight the following results: 
\begin{enumerate}
	\item The $\dot{M}_{\text{obs}}$ of Method 2 is in agreement with the $\dot{M}_{\text{obs}}$ of Method 3, thus indicating that we can place an upper limit on the $\dot{M}$ allowed by the observations of $\leq 13.5\,$M$_{\odot}$ yr$^{-1}$.
	\item The first, third and fourth methods yield consistent bolometric luminosities and predicted $\dot{M}_{\text{CF}}$; we note that Method 1 and Method 3 provide similar results for the \texttt{apec} component: in fact, by inspecting the $\chi^{2}$/D.o.f. it appears that the addition of the \texttt{mkcflow} component to Method 1 did not represent a statistically significant improvement. This conclusion can be reached also by considering the luminosities of the two components: inside the cooling region the \texttt{mkcflow} accounts for only $\approx 4.5\%$ of the observed luminosity.
	\item Considering this, our best estimate for the cooling luminosity is given by Method 1 (which is consistent with the bolometric luminosity of Method 4), thus we adopt $L_{\text{cool}} = L_{\text{bol,1}} = (1.07\pm0.06)\times10^{44}$ erg s$^{-1}$.
\end{enumerate} 
\noindent We also note that the predicted cooling rate $\dot{M}_{\text{CF}} \approx 109\,$M$_{\odot}$ yr$^{-1}$ (Method 1) overestimates the upper limit on the observed cooling rate $\dot{M}_{\text{obs}}  \lessapprox 14\,$M$_{\odot}$ yr$^{-1}$ (Method 1) by a factor of $\approx 10$. This finding reflects the cooling flow problem, and indicates that some feedback mechanism must be present in A795 to explain this difference and to prevent large amounts of gas to cool and flow to the center. We report our main hypotheses in Sect. \ref{discussion}.
\subsection{Hydrostatic mass}
\label{hydrostaticmass}
\begin{figure}
	\centering
	\includegraphics[width=\hsize]{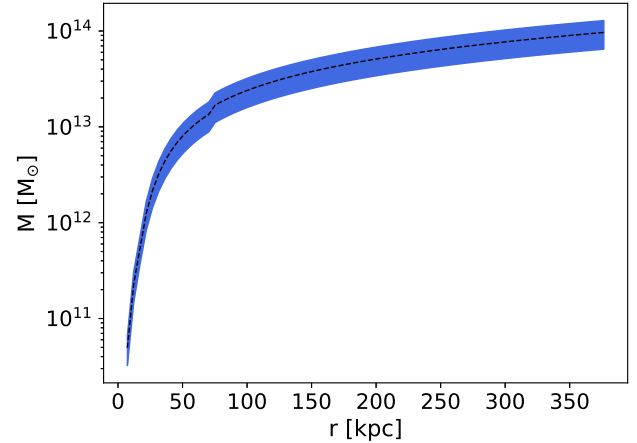}
	\caption{Hydrostatic mass profile $M_{\text{tot}}(<r)$ of A795; the blue area represents the confidence region (1$\sigma$) for the mass estimate. The discontinuity at $r\sim$70 kpc reflects the different $d\log T$/$d\log r$ inside and outside this radius.}
	\label{fig:idromass}
\end{figure}

We complemented the analysis of A795 by measuring its mass; as reported by e.g., \citet{2012AdAst2012E...6G}, one of the advantages of fitting the X-ray surface brightness profile with a $\beta$-model is that it allows to have an analytical formula for the density profile, which can be used to estimate the hydrostatic mass as: 
\begin{equation}
M_{\text{tot}}(<r) = \frac{rkT}{G\mu m_{\text{p}}}\bigg[\frac{3\beta r^{2}}{r^{2} + r^{2}_{\text{c}}} - \frac{d\log T}{d\log r}\bigg]
\label{idroeqmod}
\end{equation}
\noindent where $\beta$ and $r_{\text{c}}$ are the $\beta$-model parameters.
\\ As the best fit $\beta$-model profile of A795 does not provide a good description of the innermost point of the surface brightness profile (see Fig. \ref{fig:profsa}), we derived the hydrostatic mass for $r>$2''$\sim4.86$ kpc. In order to obtain an estimate of the temperature gradient in A795, we fitted the temperature profile of the deprojected spectral analysis (Subsect. \ref{spectralA795}) in the log$-$log space with two power-laws, one describing the temperature drop in the inner $\sim$70 kpc and the other one accounting for the flatter profile at larger radii (see Fig. \ref{fig:deprojtempab}); the best fit linear regression yields $\log T = (0.13 \pm 0.01)\log r + (4.74\pm0.02)$ within $\sim$70 kpc and $\log T = (-0.02\pm0.01)\log r + (8.29\pm1.61)$ outside. By substituting in Eq. \ref{idroeqmod} the slope of the two power-laws and the values of $\beta$ and $r_{\text{c}}$ from the best $\beta$-model fit to the surface brightness profile of A795, we produced the hydrostatic mass profile of Fig. \ref{fig:idromass}. 
\\\citet{2006MNRAS.369.2013R} reported that the use of $\beta$-model for the hydrostatic mass estimates lead to systematic uncertainties of $\sim$20\% with respect to the true mass estimate; moreover, residual gas motions in the central regions of galaxy clusters (such as sloshing) represent an additional bias in the cluster mass estimate of $\approx$6-9 \% (e.g., \citealt{2009ApJ...705.1129L,2020MNRAS.495..864A}). Therefore, beside propagating the errors on $kT$, $\beta$, $r_{\text{c}}$ and $\alpha$, we added a 30$\%$ error to account for the approximation of using hydrostatic equilibrium to describe a disturbed system.
\\ Outside \textit{Chandra}'s field of view, we obtained the mass profile of this cluster by fitting a NFW profile \citep{1996ApJ...462..563N} to the hydrostatic mass\footnote{Fitting was performed using the \texttt{optimize.curve\_fit} tool of \textit{Python3.5.4}}. Subsequently, we extrapolated the fitted profile to $r_{500}$ (the radius at which the enclosed mean density is 500 times the critical density of the Universe at A795's redshift) and to the virial radius $r_{\text{vir}} = r_{200}$. We measured $r_{500}=732\pm85\,\text{kpc}$ and $r_{\text{vir}}=$1.0$\pm$0.2 Mpc; a concentration c$_{500}=2.4\pm0.2$ and c$_{\text{vir}} = 3.3\pm0.3$; a total mass $M_{500}=2.3\pm1.0\times10^{14}$M$_{\odot}$ at $r_{500}$ and $M_{\text{vir}}$ = 3.1$\pm$1.4$\times10^{14}$ M$_{\odot}$ at $r_{\text{vir}}$ (consistent with the values reported in \citealt{2013ApJ...767...15R}).
\subsection{Cold fronts and sloshing}
\label{specdis}
\begin{figure}
	\centering
	\includegraphics[width=\hsize]{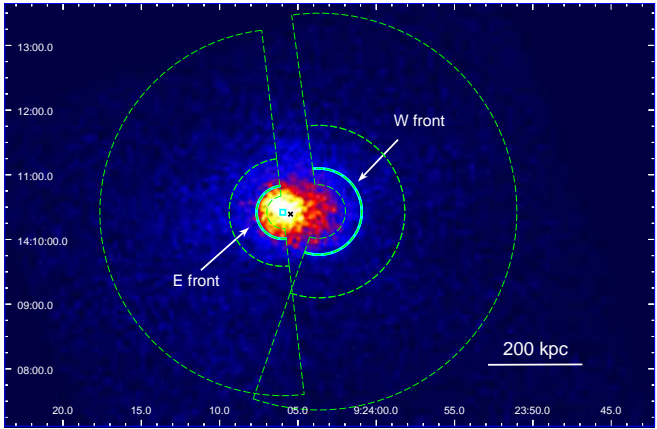} 
	\caption{A795 0.5-7 keV image, Gaussian-smoothed with a kernel radius of 5'', showing the sectors (green, dashed regions) used to study the thermodynamical properties of the two fronts (cyan arcs). The cyan box and the black cross mark the position of the X-ray peak and of the central AGN, respectively.}
	\label{fig:sectors}
\end{figure}

\begin{figure*}
	\centering
	\subfloat[]{\includegraphics[width=0.25\hsize]{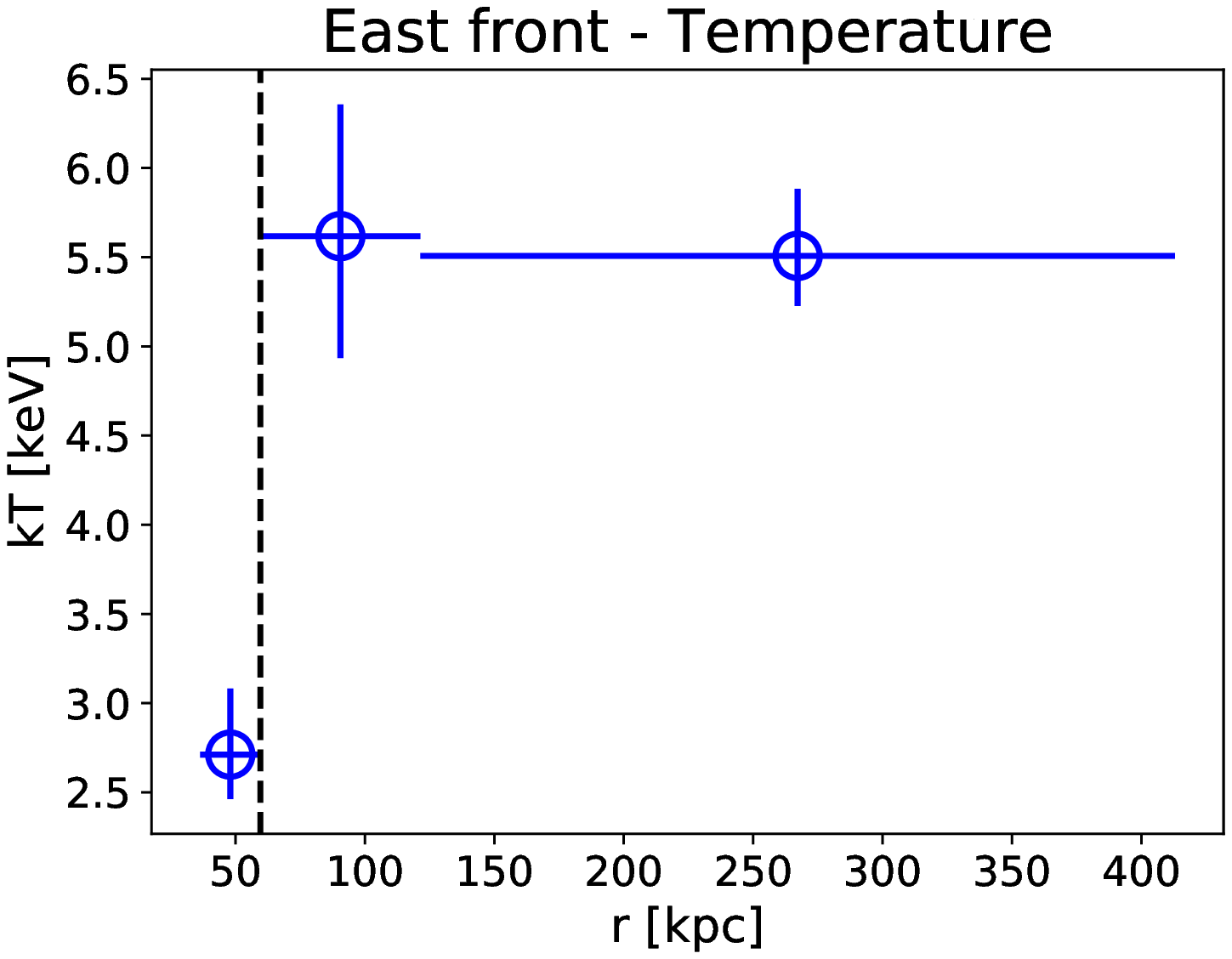}} 
	\subfloat[]{\includegraphics[width=0.25\hsize]{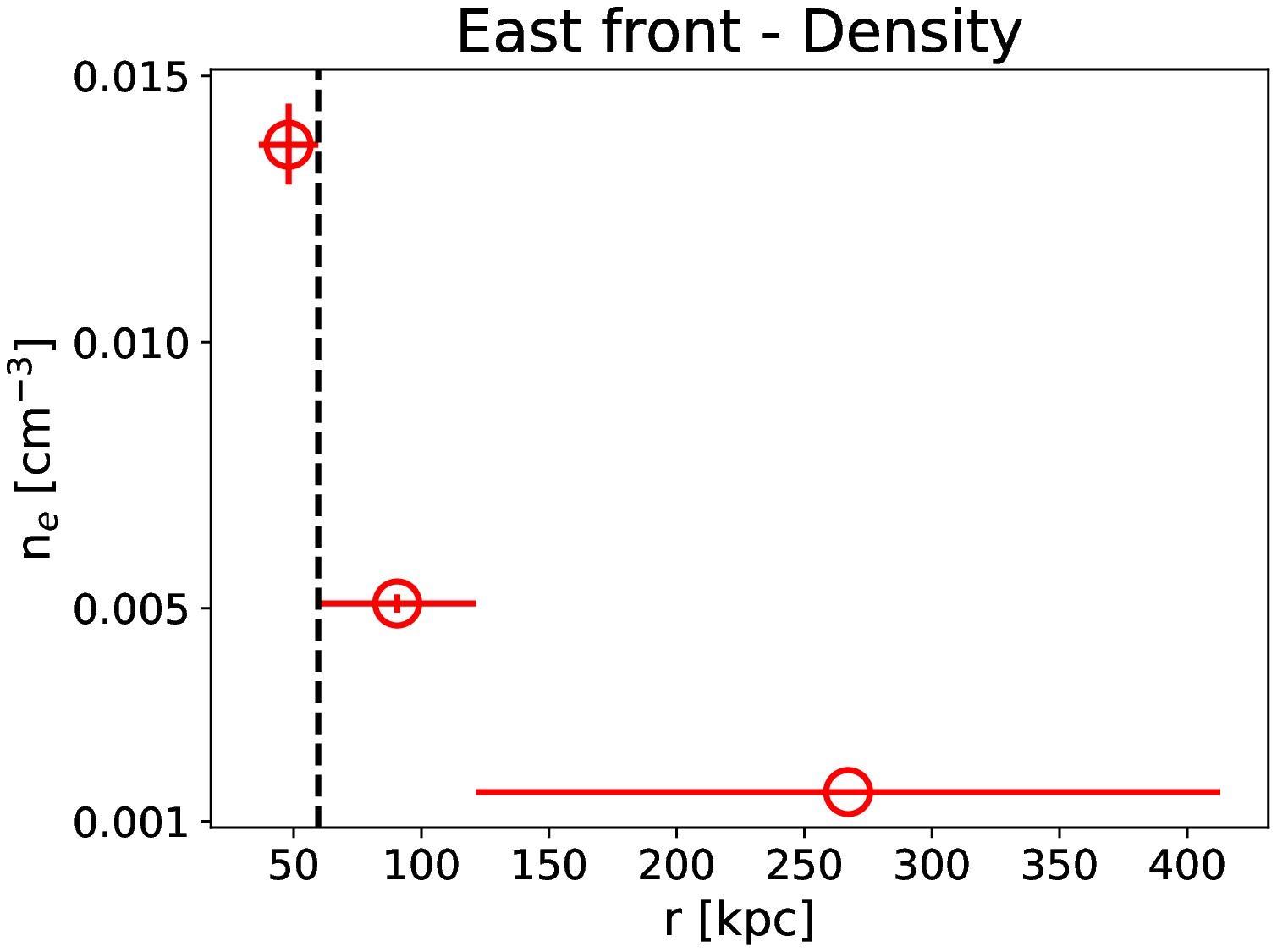}} 
	\subfloat[]{\includegraphics[width=0.25\hsize]{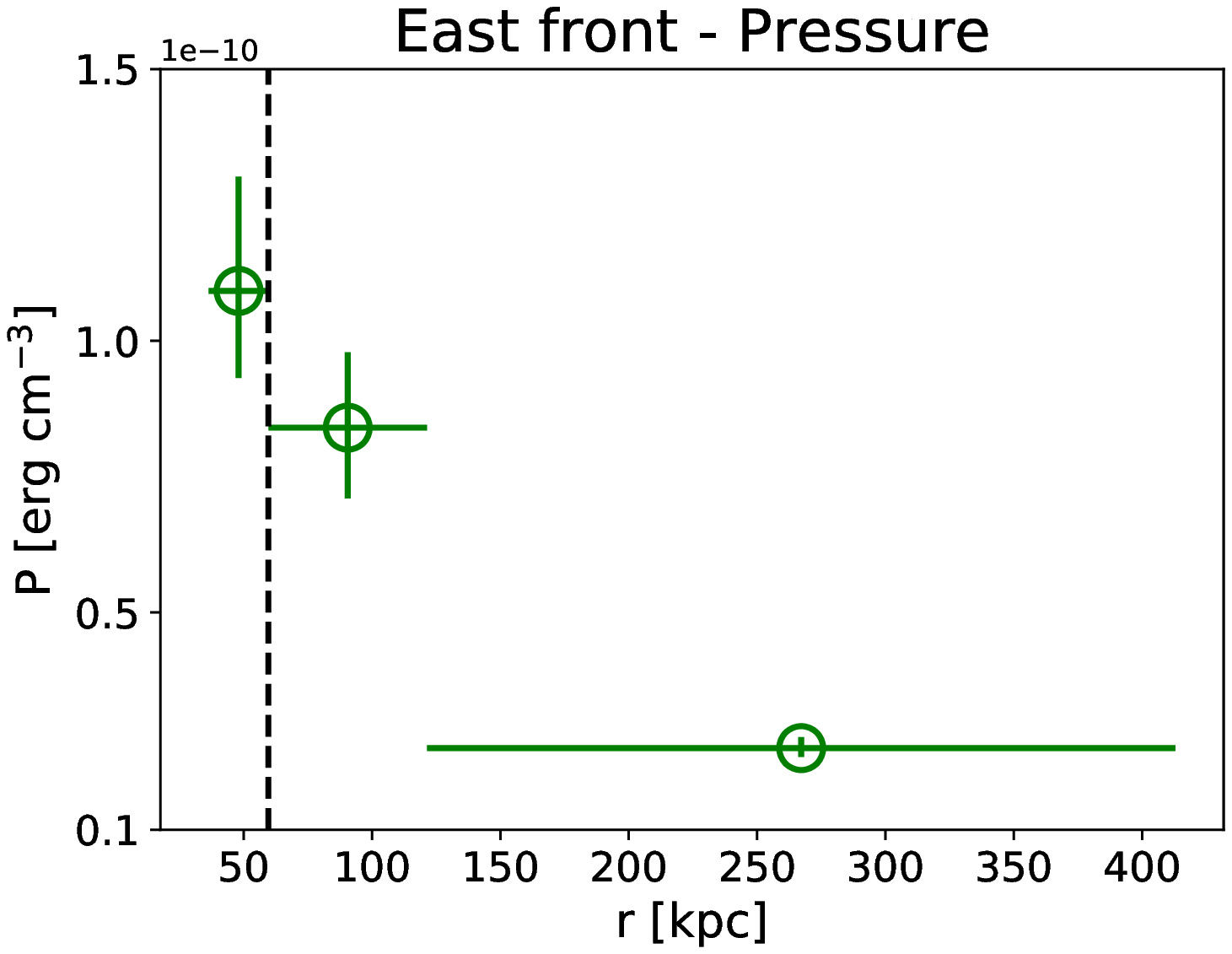}} 
	\subfloat[]{\includegraphics[width=0.25\hsize]{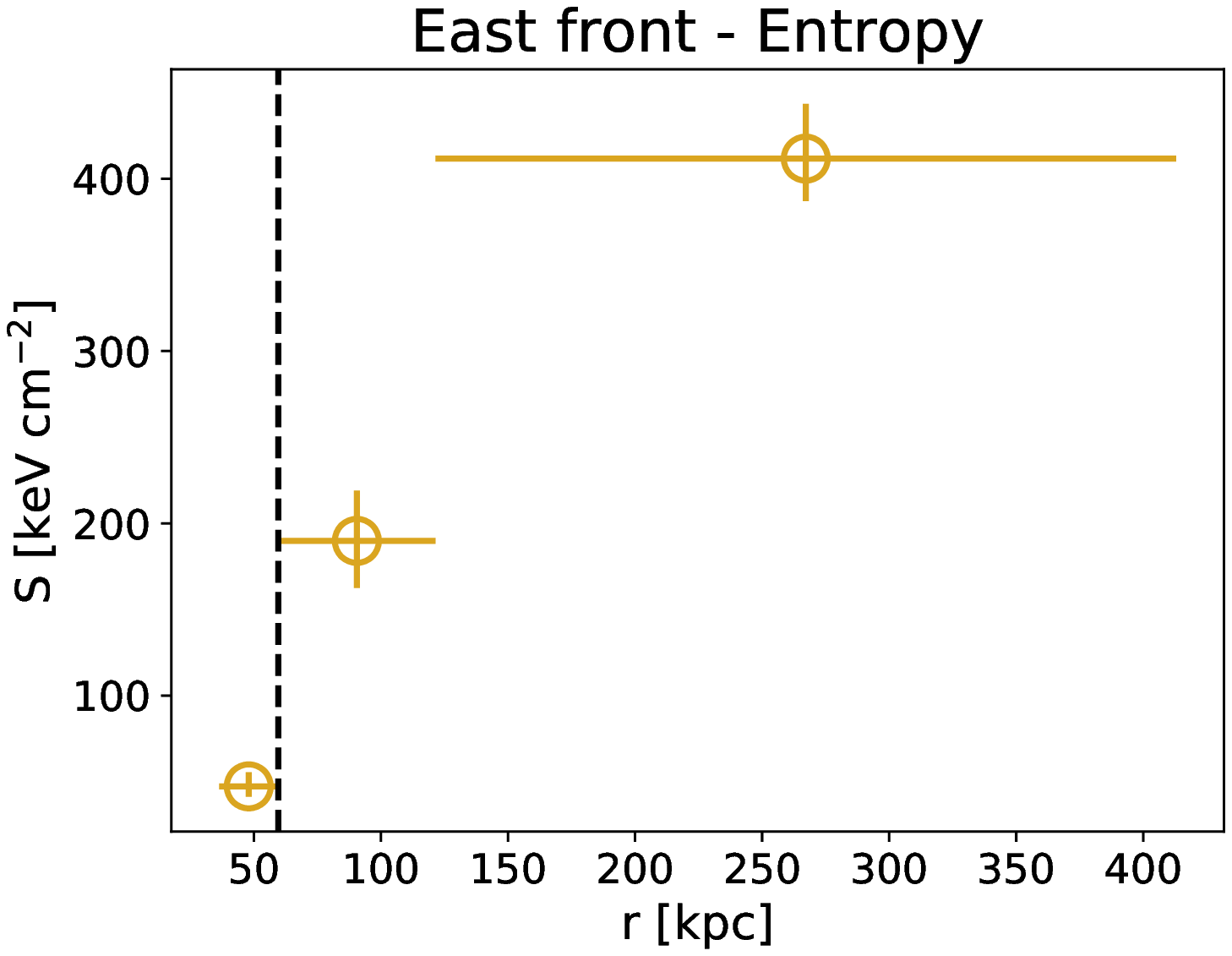}} \\
	\subfloat[]{\includegraphics[width=0.25\hsize]{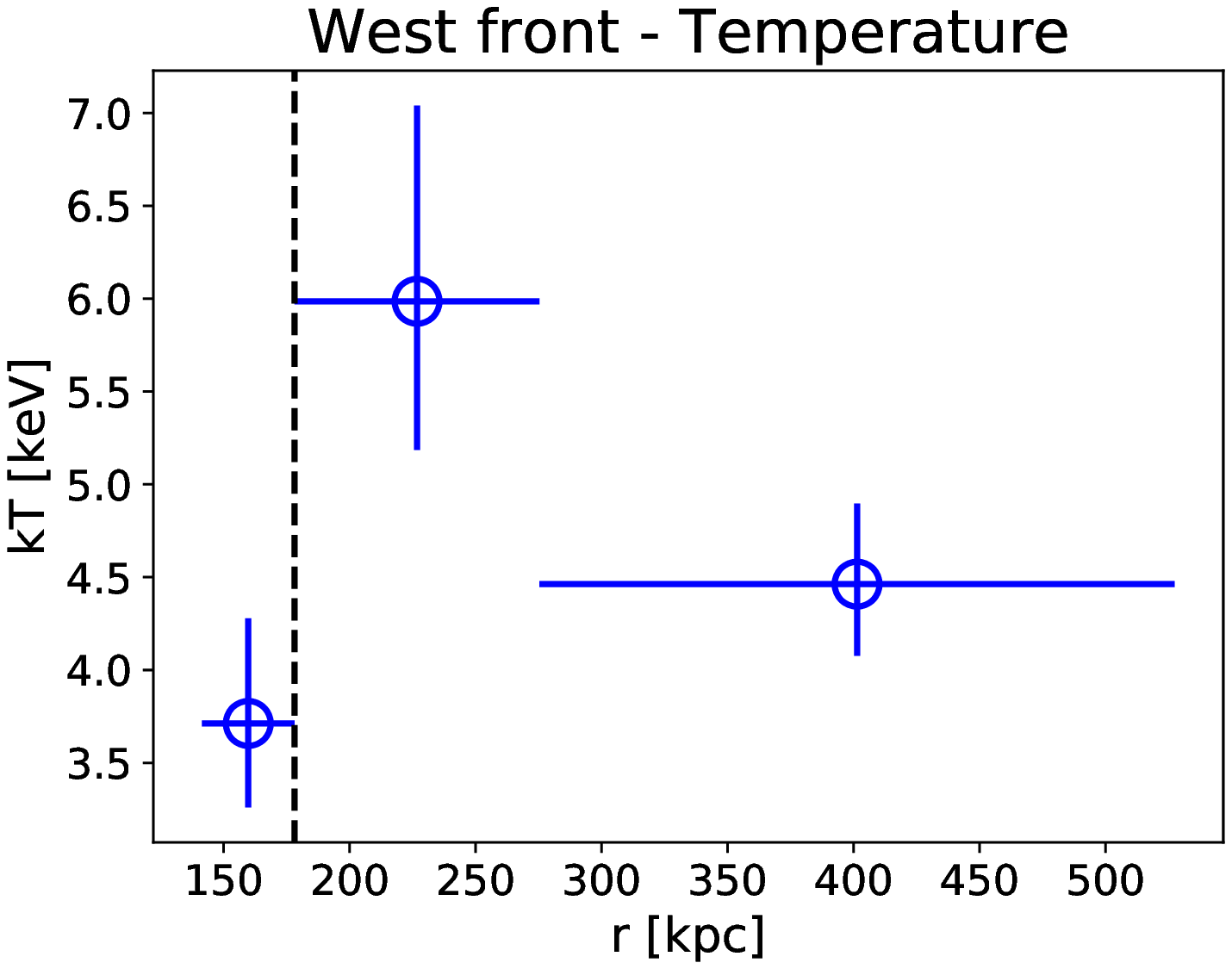}} 
	\subfloat[]{\includegraphics[width=0.25\hsize]{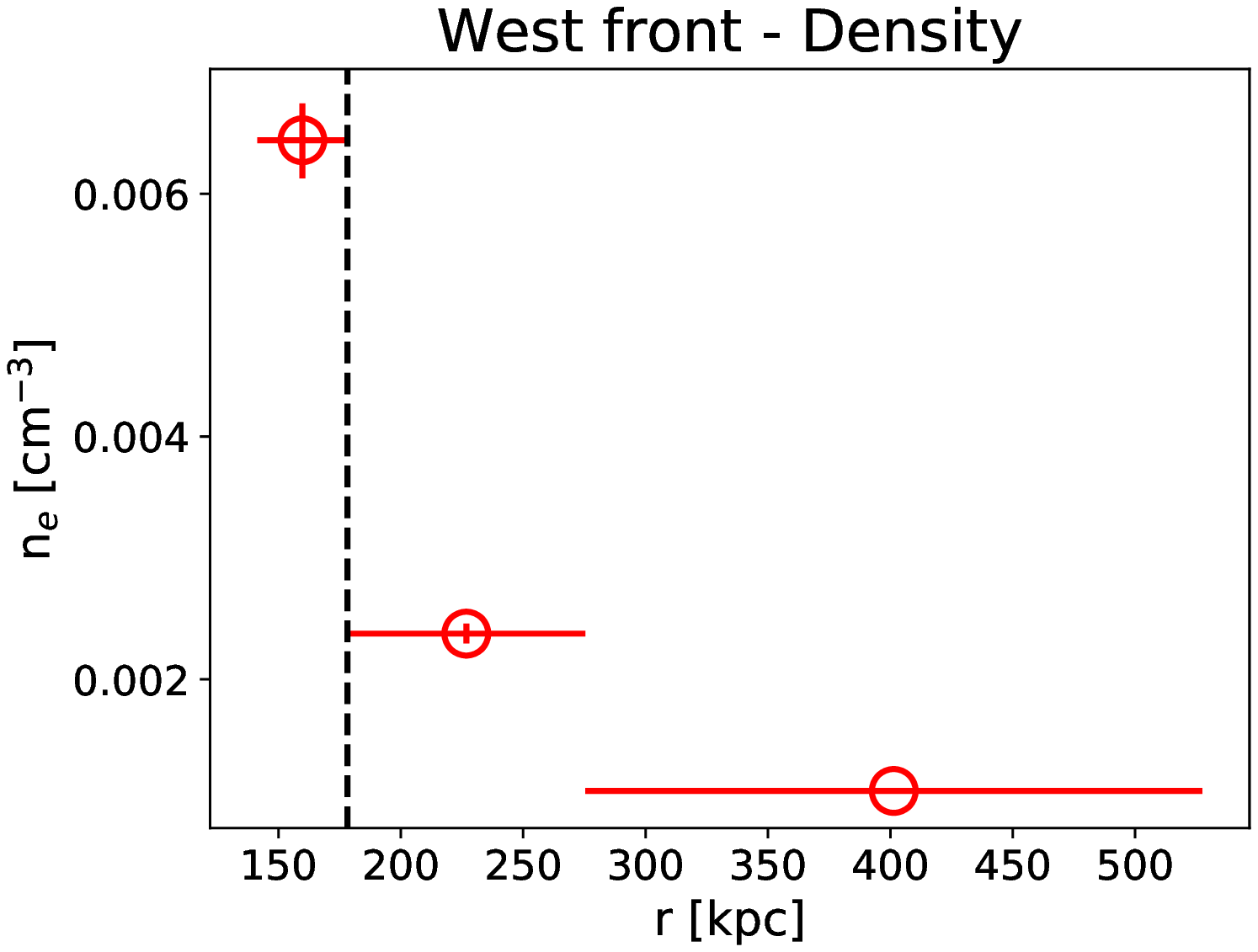}} 
	\subfloat[]{\includegraphics[width=0.25\hsize]{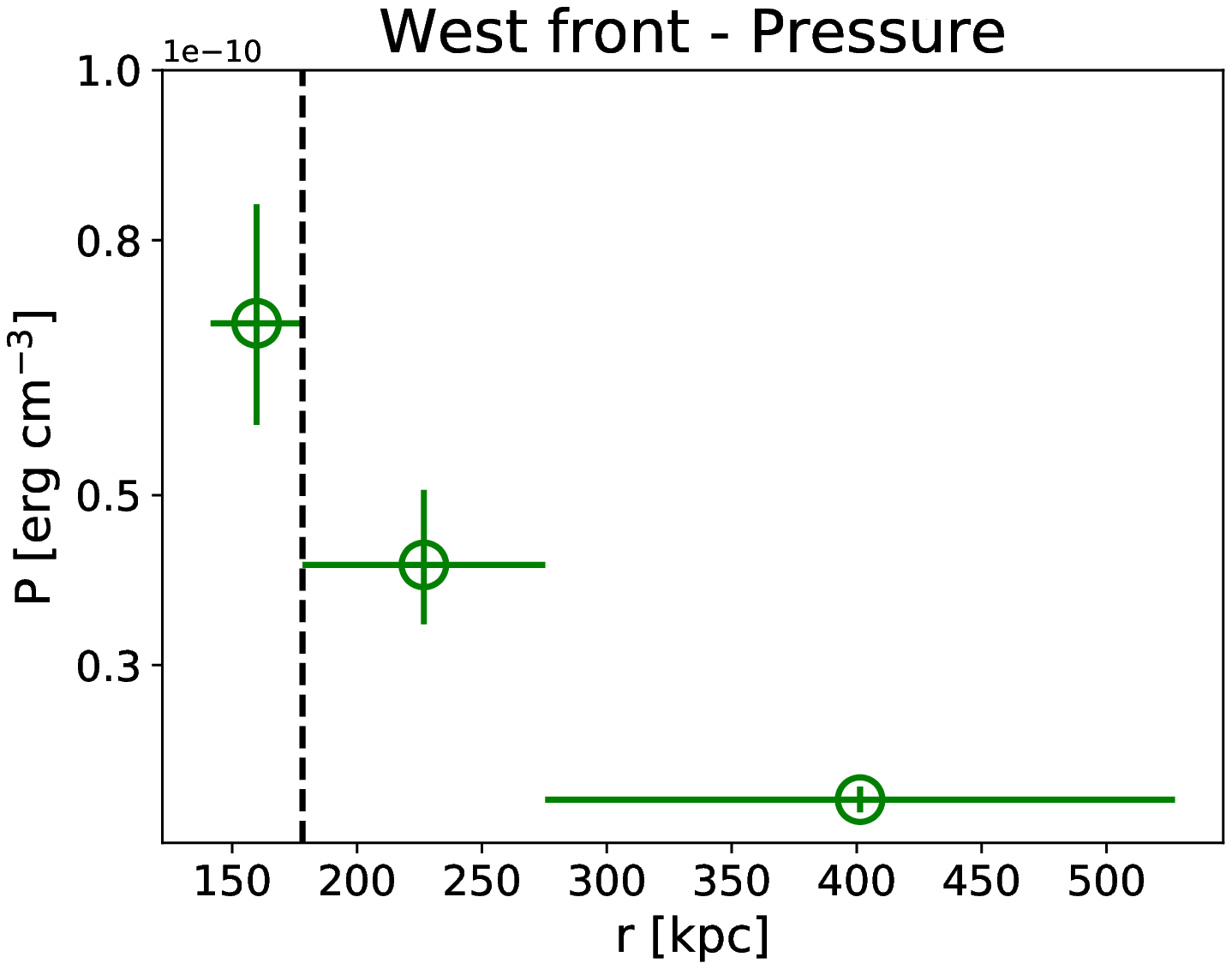}}
	\subfloat[]{\includegraphics[width=0.25\hsize]{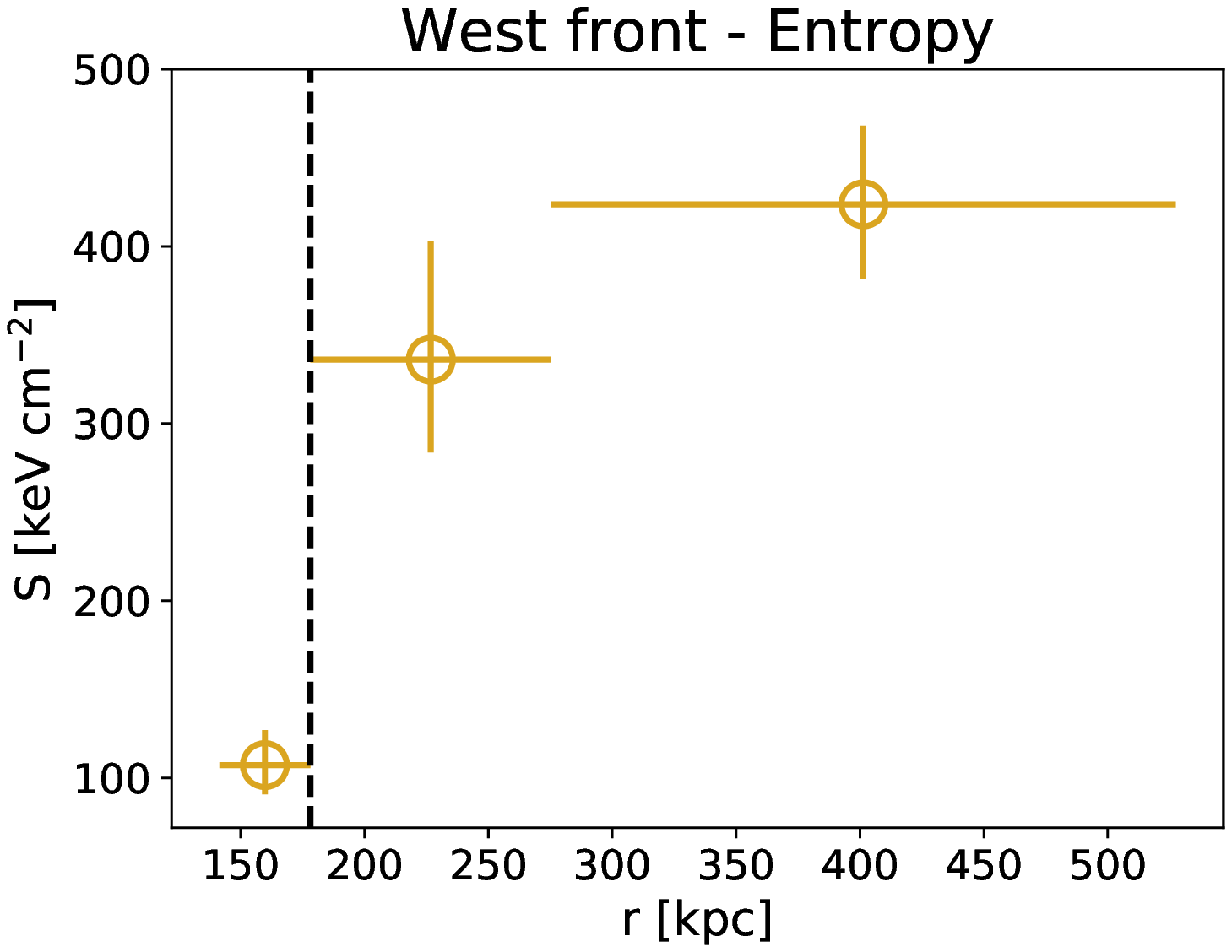}} 
	\caption{Temperature, density, pressure and entropy profiles across the east front (\textit{top row}), and the west front (\textit{bottom row}). The x-axis of the plots indicates the distance of each sector from the X-ray peak (while the sectors used for the analysis of the west front are centered at $\approx$80 kpc from the X-ray peak). The black vertical line represents the position of the cold fronts (determined from the surface brightness analysis of Sect. \ref{morphoA795}.}
	\label{fig:ewprofili}
\end{figure*}

In order to spectroscopically confirm the hypothesis on the cold front nature of the two surface brightness jumps in the ICM of A795, we measured the thermodynamic properties of the ICM in sectors along each discontinuity (Fig. \ref{fig:sectors}). The width of the pre-front and post-front bins has been selected to have a sufficient statistics while avoiding any contamination from different phases of the ICM. We also added an external bin (reaching the edge of the chip) to account for projection effects. 
\begin{table*}
		\centering
		\caption{Spectral analysis of the two cold fronts: (1) name of the edge and distance from the X-ray peak; (2) inner and outer radius of the sector; (3) net photon counts (fraction w.r.t. the total counts from the same region); (4) temperature; (5) electron density; (6) pressure of the ICM; (7) entropy of the ICM. The $\chi^{2}$/D.o.f. is 344.8/388 (0.89) for the east front and 364.1/377 (0.97) for the west front.}
		\label{tab:propertiesew}
		\begin{tabular}{ccccccc}
			\hline
			Discontinuity & R$_{\text{i}}$ - R$_{\text{o}}$ & Counts& $kT$ &   $n_{\text{e}}$ &$P_{\text{ICM}}$ & $K_{\text{ICM}}$\\
			&kpc & & keV &  10$^{-3}$ cm$^{-3}$ & 10$^{-11}$ erg cm$^{-3}$ & keV cm$^{2}$\\
			\hline
			\rule{0pt}{2ex}\multirow{3}{*}{E front (59.7 kpc)}& 36.4 - 59.7 &1660 (98.2 \%) & 2.71$^{+0.37}_{-0.25}$  &  13.7$^{+0.8}_{-0.7}$& 10.9$^{+2.1}_{-1.6}$ & 47.3$^{+8.2}_{-6.0}$\\
			& 59.7 - 121.5  &2561 (94.0 \%) & 5.62$^{+0.74}_{-0.68}$  &  5.1$^{+0.2}_{-0.2}$& 8.4$^{+1.4}_{-1.3}$ & 189.8$^{+29.2}_{-27.3}$\\
			& 121.5 - 412.9 & 7704 (77.8 \%)   & 5.51$^{+0.38}_{-0.28}$  &  1.5$^{+0.1}_{-0.1}$& 2.5$^{+0.2}_{-0.2}$ & 411.7$^{+31.7}_{-24.6}$ \\
			\hline
			\rule{0pt}{2ex}\multirow{3}{*}{W front (177.8 kpc)}& 60.4 - 97.2 &1593 (94.9 \%) & 3.71$^{+0.57}_{-0.45}$  & 6.4$^{+0.3}_{-0.3}$& 7.0$^{+1.4}_{-1.2}$ & 107.2$^{+19.7}_{-16.5}$\\
			&97.2 - 194.3 &2570 (85.0 \%) & 5.99$^{+1.05}_{-0.80}$   & 2.4$^{+0.1}_{-0.1}$&4.2$^{+0.9}_{-0.7}$ &336.1$^{+67.0}_{-52.5}$\\
			&194.3 - 446.4 &4257 (61.6 \%)   & 4.46$^{+0.43}_{-0.39}$  & 1.1$^{+0.1}_{-0.1}$&1.4$^{+0.2}_{-0.2}$ &423.7$^{+44.5}_{-41.5}$\\
			\hline
		\end{tabular}
\end{table*}	
\\Spectra were fitted using a \texttt{projct$\ast$tbabs$\ast$apec} model with $kT$, $Z$ and $norm$ parameters of the \texttt{apec} component free to vary, while the column density and redshift were fixed. We used Eq. \ref{nenorm} to obtain the density of the ICM within each sector, and computed the pressure $P_{\text{ICM}}$ and the entropy $K_{\text{ICM}}$ across the two fronts. Tab. \ref{tab:propertiesew} lists the best fit parameters and the thermodynamical properties of the two discontinuities, while Fig. \ref{fig:ewprofili} shows the temperature, density, pressure and entropy profiles along the two discontinuities; we emphasize the following results:
\begin{enumerate}
	\item \textbf{E front:} The temperature profile (Fig. \ref{fig:ewprofili} panel \textit{a}) confirms that this edge is indeed a cold front, since the inner side of the front is colder than the outer side; the pressure profile is continuous and the entropy profile shows a prominent jump. The temperature ratio is $T_{\text{out}}/T_{\text{in}} = 2.07 \pm 0.53$ ($\sim$2$\sigma$).
	\item \textbf{W front:} The thermodynamical properties of this edge of Fig.\ref{fig:ewprofili} (bottom row) are also typical of cold fronts. This front is not as sharp as the eastern one, as the temperature ratio is $T_{\text{out}}/T_{\text{in}} =1.61 \pm 0.46$ ($\sim$1.3$\sigma$).
\end{enumerate} 
\indent The spectral analysis has then confirmed that the ICM of A795 has been perturbed, and sloshing has displaced the cool gas from the center, creating a spiral morphology along which we detected two cold fronts. We determine the timescale that regulates the ICM oscillation in Subsect. \ref{sloshingtime}. 

\subsection{The FR0 J092405.30+14 and the surrounding environment}
\label{J092405.30+14}
\begin{figure}
	\centering
	\subfloat[]{\includegraphics[width=\hsize]{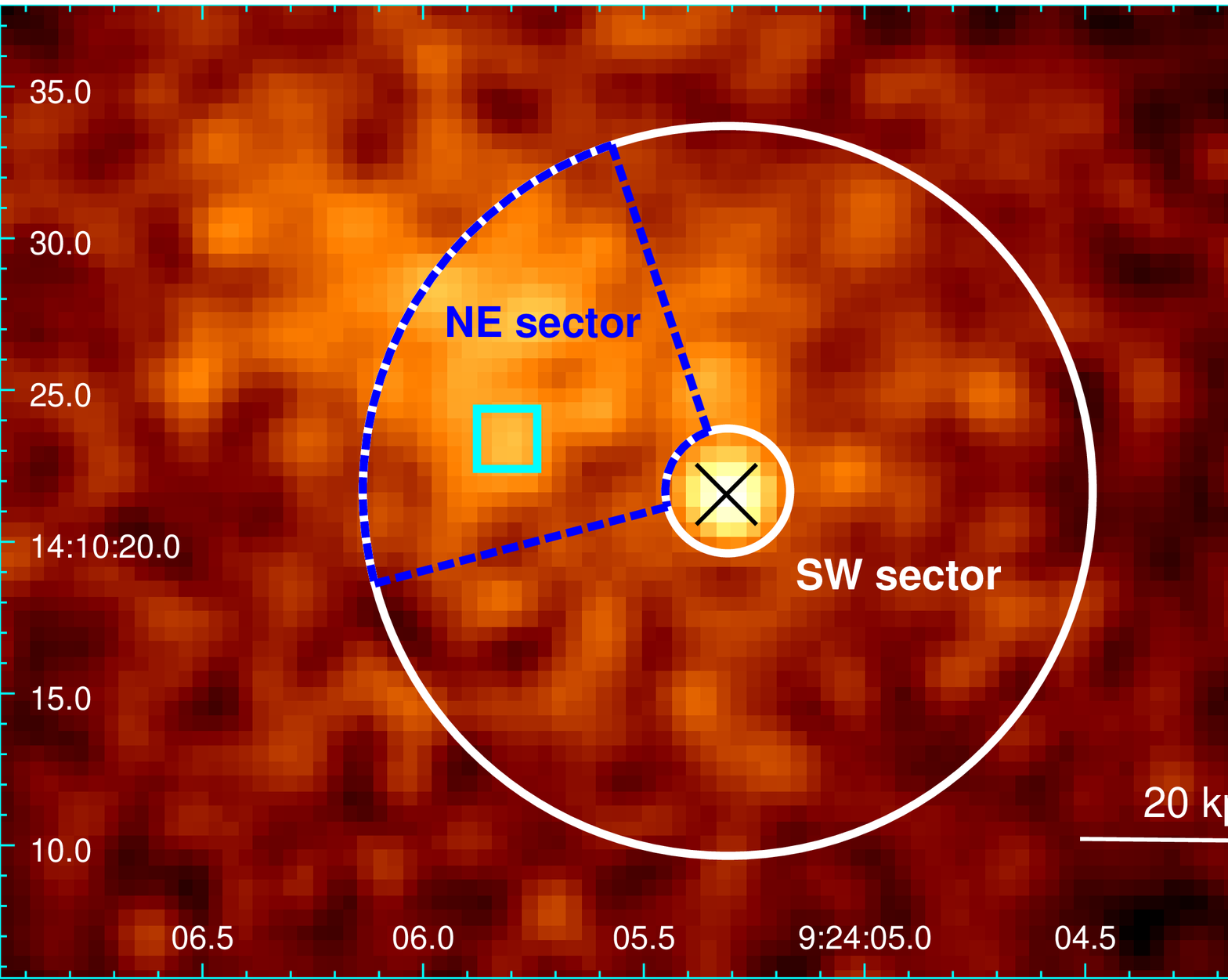}} 
	\caption{
	0.5-7 keV image of A795, Gaussian-smoothed with a kernel radius of 1.5''; the regions used to study the multi-phase gas around J092405.30+14 are overplotted: the white annulus is the extraction region for the ambient gas spectrum; the blue dashed sector shows the NE sector, while the SW sector is composed by the remaining portion of the annulus. The cyan box and the black cross denote the positions of the X-ray peak and the BCG, respectively.}
	\label{fig:agnregion}
\end{figure}

We extracted the spectrum of the central AGN in the 0.5-7 keV band from the same region excluded during the spectral analysis of the cluster (an ellipse centered onto the BCG with semi-major axis 1.9'', semi-minor axis 1.7'' and position angle 147.4$^{\circ}$); the background spectrum has been extracted from the blanksky event file in the same region\footnote{We verified that considered the poor statistics (145 counts before background subtraction) the use of blanksky or of local background (an annulus surrounding the source) provides consistent results. Since the local background around the source is patchy and filamentary (see Fig. \ref{fig:agnregion}) we believe that the most conservative choice is to use the blanksky event file.}. 
\\The source spectrum has 88 net counts in the 0.5-7 keV band; we decided to group the data with to obtain at least 1 count per bin and to use the Cash statistics. We fitted the data with a \texttt{tbabs$\ast$po} model, fixing the $N_{\text{H}}$ to the Galactic value and leaving the normalization $K$ and the photon index $\Gamma$ of the power-law free to vary. Tab. \ref{tab:tbabspoapec} lists the best fit parameters of this model: we measured an X-ray flux of $F_{\text{2-10 keV}} = 1.8^{+0.4}_{-0.3} \times 10^{-14}$ erg cm$^{-2}$ s$^{-1}$, and a corresponding luminosity of $L_{\text{2-10 keV}} = 9.1^{+1.8}_{-1.6} \times 10^{41}$ erg s$^{-1}$. By comparing a soft (0.5-4 keV) and a hard (4-7 keV) image of J092405.30+14 we verified that the AGN fades into the background in the hard image, which is in good agreement with the steep value of the photon index ($\Gamma = 2.13^{+0.16}_{-0.15}$). Our results are in agreement with the analysis of \citet{2018MNRAS.476.5535T}. In addition to this, considering our updated measurement of the X-ray luminosity the source still lies within the correlation between $L_{\text{X}}$ and $L_{\text{5 GHz}}$ presented in Fig. 4 (lower panel) of \citet{2018MNRAS.476.5535T}, i.e. the 2-10 keV emission is probably of non-thermal origin.
\begin{table}
	\centering
	\caption{Spectral analysis of the central AGN: (1) spectral index of the power-law; (2) spectrum normalization, in units of $10^{-6}$ photons keV$^{-1}$ cm$^{-2}$ s$^{-1}$; (3) Cash statistics/degrees of freedom; (4) 2-10 keV flux, in units of $10^{-14}$ erg cm$^{-2}$ s$^{-1}$; (5) 2-10 keV luminosity, in units of $10^{41}$ erg s$^{-1}$.}
	\label{tab:tbabspoapec}
	\begin{tabular}{ccccc}
		\hline
		 $\Gamma$ & $K$ & C/D.o.f. & $F_{\text{2-10 keV}}$ & $L_{\text{2-10 keV}}$\\
		\hline
		   $2.13^{+0.16}_{-0.15}$ & $8.32^{+0.78}_{-0.74} $& 85/91 & $1.8^{+0.4}_{-0.3}$ & $9.1^{+1.8}_{-1.6}$\\
		\hline
	\end{tabular}
\end{table}	
\\ \noindent We found a positive residual around $\sim 0.9$ keV: we added an \texttt{apec} component to the absorbed power-law model, and found a slight improvement in the fit for a thermal plasma with temperature kT$\sim$1.2 keV, suggesting a possible thermal contribution from the hot corona of the elliptical host galaxy to the source spectrum. We also excluded that intrinsic absorption is present: there are no negative residuals below $\approx$1 keV, and the photon index did not assume very flat values, which would suggest the presence of an intrinsic absorber (e.g., \citealt{2020MNRAS.493.4355M}). Adding an intrinsic absorption component returned only an upper limit for $N_{\text{H}}$ of $<0.5\times10^{22}$ cm$^{-2}$: such a low value is similar to those found in FRI-LERGs (e.g., \citealt{2006A&A...451...35B,2008A&A...489..989B}), supporting the hypothesis that also the nuclear region of FR0s is not filled with cold, dense matter \citep{2018MNRAS.476.5535T}. 
\\ To investigate the properties of the ambient gas, an annulus of inner radius 2'' and outer radius 12'', centered on the BCG (see Fig. \ref{fig:agnregion}), has been used to extract the spectrum of the ICM surrounding J092405.30+14. We extracted the background spectrum from the blanksky event file using the same region, and fitted the data in the 0.5 - 7 keV with a \texttt{tbabs$\ast$apec} model, using the $\chi$-statistics. Results are presented in Tab. \ref{tab:ambienticm}. 
\\ Considering the sloshing motion of the ICM and the geometry of the cold gas spiral, we expected that the north-east side of the AGN, in the proximity of the X-ray peak, could consists of lower temperature gas w.r.t. the south-west side. We splitted the annulus in two sectors, the first one including the X-ray peak, and the second one the remaining portion of the region (NE and SW sectors in Fig. \ref{fig:agnregion}). By fitting the spectra with another thermal model (Tab. \ref{tab:ambienticm} second and third row), we confirmed that the ICM around the FR0 is multiphase, and subject to the temperature gradients induced by sloshing. 
\begin{table}
\centering
	\caption{Spectral analysis of the ICM around J092405.30+14: (1) considered region (see Fig. \ref{fig:agnregion}); (2) temperature; (3) metallicity; (4) normalization of the spectrum; (5) $\chi^{2}$/degrees of freedom.}
	\label{tab:ambienticm}
		\begin{tabular}{lcccc}
			\hline
			Region &  $kT$ & $Z$ & $norm$& $\chi^{2}$/D.o.f. \\
					    & keV & Z$_{\odot}$ & & \\
			\hline
			Annulus & 3.81$^{+0.23}_{-0.23}$  & 0.75$^{+0.20}_{-0.18}$ & 5.26$^{+0.26}_{-0.26}$ & 77.1/70 \\
			NE-sector& 3.54$^{+0.41}_{-0.33}$  & 0.97$^{+0.51}_{-0.38}$ & 1.74$^{+0.19}_{-0.19}$ & 30.0/26 \\
			SW-sector & 4.30$^{+0.30}_{-0.29}$  & 0.86$^{+0.30}_{-0.26}$ & 3.34$^{+0.21}_{-0.21}$ & 45.7/49 \\
			\hline
		\end{tabular}
\end{table}	
\\ The innermost annulus of the (deprojected) radial spectral analysis overlaps with the annulus surrounding the AGN, therefore we assumed the density of the ICM around J092405.30+14 to be equal to the value measured in Sect. \ref{spectralA795} of $n_{\text{e}} = (2.14\pm0.11) \times 10^{-2}$ cm$^{-3}$.
\section{Discussion}
\label{discussion}
In this section we perform an inside-out discussion of the properties we measured for the environment of A795: we start by arguing the possible link between the central FR0's radio size and the surrounding ambient (kpc scale); we proceed to investigate whether and how AGN feedback is acting in this cluster (tens of kpc), and we estimate the age of the large scale (hundreds of kpc) sloshing spiral. At last, we study the properties of the cluster-scale radio source we derived by inspecting survey radio data.
\subsection{Is the ICM preventing the jet expansion in J092405.30+14?}
As proposed by \citet{2015A&A...576A..38B}, a high density gas might be capable of decelerating and disrupting the radio jets. Here we discuss both the density of the surrounding medium, and the effects of sloshing-induced turbulence of the jet stability. If the cluster's gas exhibited extreme thermodynamical conditions (i.e. peculiar with respect to those of the ICM surrounding extended FRIs in BCGs) we could infer that the environment of J092405.30+14 is interfering with the jet propagation.
\begin{enumerate}
	\item \textbf{The density of the ICM:} The electron density of the ICM around J092405.30+14 is $2.14 \times 10^{-2}$ cm$^{-3}$. This value is in line with typical ICM densities around FRIs at the center of galaxy clusters ($\approx 10^{-3} - 10^{-1}$ cm$^{-3}$, e.g., \citealt{2010ApJ...713.1037H,2017ApJ...843...28M}): this suggests that if FR0s' compactness is due to the frustration and disruption of the jet on small scales ($\approx 1 - 15$ kpc), the local density of the ICM is not responsible for it, at least in A795.
	\item \textbf{The effect of sloshing:} In Subsect. \ref{J092405.30+14} we found that there is a multiphase medium around J092405.30+14, suggesting that even the central regions ($r \lessapprox 30\,\text{kpc}$) are affected by the large scale oscillation. Therefore, we discuss the possibility that the ICM turbulence is affecting the stability of radio jets. In clusters and groups displaying sloshing motions of the gas there are FRIs whose jets have not been disrupted (e.g., the FRI at the center of the galaxy group 3C449, \citealt{2013ApJ...764...83L}; the radio galaxy in the NGC 1550 group, \citealt{2020MNRAS.496.1471K}). Thus, we argue that unless there are intrinsic differences between the jets of FR0s and FRIs, sloshing alone is not capable of preventing their propagation: only if the jet presents a low Lorentz factor and is launched by a slowly spinning black hole, then turbulence induced by the gas motion might quench the weak jet as soon as it enters the ICM (e.g., \citealt{2019MNRAS.482.2294B}).
	\item \textbf{Warm ionized gas:} The ICM is not the only turbulent gas phase found in the proximity of J092405.30+14: \citet{2016MNRAS.460.1758H} studied the warm ionized gas dynamics in the cores of 73 galaxy clusters. They found that a roughly spherical cloud of H$\alpha$-emitting gas with an extent of $\sim$11 kpc surrounds the BCG of A795; the high FWHM of the H$\alpha$ line ($400-800$ km/s) might suggest that we are seeing a system of H$\alpha$ clouds along the line of sight, with high-velocity random motions. It is possible that the warm gas around J092405.30+14 might be slowing the jet expansion in the outer medium. To verify this hypothesis would require to model the clumpiness of these gas clouds, so that the jet frustration could also be compatible with the lack of X-ray absorbing cold gas suggested by the spectral modeling (Subsect. \ref{J092405.30+14}). If future observations of A795 will detect molecular gas clouds, co-spatial with the warm phase, it might become possible to put stronger constraints on the jet frustration. Once again, however, it should be noted that H$\alpha$-emitting clouds have also been detected by \citet{2016MNRAS.460.1758H} around FRIs at the center of galaxy clusters: this suggests that the jet has to be intrinsically weak to get easily disrupted by the warm and/or cold gas. \\As a note of caution, we point out that the H$\alpha$ line broadening might be caused by non-gravitational kinematics, for example outflow motions (see e.g., \citealt{2017ApJ...845..131K}): the warm gas velocity dispersion is a factor of $\gtrapprox 2$ greater than the stellar velocity dispersion ($\sigma_{\star}$ = 261$\pm$9 km s$^{-1}$, \citealt{2015ApJS..219...12A}), implying that the hydrogen gas might not only be tracing the gravitational potential of the host galaxy, but could also be influenced by an additional non-virial component. 
\end{enumerate} 
\indent We argue that even the co-existence of sloshing and H$\alpha$ is likely not able to explain the compactness of the radio source: it is possible to find examples of galaxy clusters displaying sloshing motions with H$\alpha$ clouds in their cores which still host extended FRIs at their center (e.g., A3581, \citealt{2013MNRAS.435.1108C}; A2495, \citealt{2019ApJ...885..111P}; A1668, \citealt{pasini2021}). \\The overall considerations point to the conclusion that the reason(s) behind FR0s' radio compactness has to be searched primarily in the intrinsic properties of the jet and/or BH spin. The environment can still play a role, although not the major one. In fact, our X-ray study of the environment of J092405.30+14 has revealed a complex, multiphase and possibly turbulent ambient, but in line with typical properties of the ICM surrounding extended radio galaxies. This might highlight the critical role of the jet stability, and - in turn - of the black hole spin: unless the central engine of this FR0 has peculiar parameters that lead to the formation of unstable jets, turbulence alone cannot explain the radio compactness.
\subsection{AGN feedback in A795}
\begin{figure}
	\centering
	\includegraphics[width=\hsize]{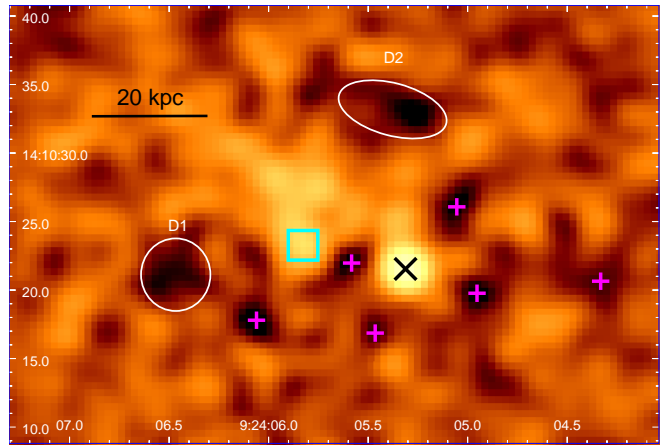}
	\caption{A795 0.5-2 keV unsharp mask image, obtained by subtracting two images smoothed with a Gaussian of 1'' and 5'' axes, respectively; the positions of the AGN (black cross), of the X-ray peak (cyan box), and of the two depressions are marked. The pink plus signs are examples of the depressions with a negligible deviations (2-8$\%$) w.r.t. the best $\beta$ model.}
	\label{fig:cavpos}
\end{figure}

In this subsection we investigate the AGN feedback mechanisms that might be inhibiting cooling in A795, in particular by looking for the presence of cavities in the ICM. After testing several smoothing combinations, by subtracting two images of the cluster (smoothed with a Gaussian of 1'' and 5'' axes, respectively) we obtain the unsharp mask image that better emphasizes substructures in the ICM (Fig. \ref{fig:cavpos}): by inspecting the result, we identify two depressions (named D1 and D2) close to the BCG and symmetric with respect to the X-ray peak.
\\ We note that the other depressions visible in Fig. \ref{fig:cavpos} (marked with pink circles) do not represent a significant deviation w.r.t. the best $\beta$ model fit (between 2-8$\%$, with a significance of 1.04$\sigma$) so we do not include them in our analysis. On the contrary, D1 and D2 show a deficit of $\approx34 \%$ and $\approx21 \%$ respectively, which is typical of X-ray cavities (e.g., \citealt{2007ARA&A..45..117M}). The cavity significance is 2$\sigma$ for D1 and at 1.4$\sigma$ for D2; according to the classification of \citet{2015ApJ...805...35H}, the significance between 1-3$\sigma$ of our depressions indicates that deeper observations are necessary to obtain a clear cavity detection; thus, we refer to D1 and D2 as \textit{putative} cavities. \\ Under the assumption that the two X-ray depressions are real cavities, it is possible to obtain the cavity power $P_{\text{cav}} = 4pV/$t$_{\text{age}}$, where $p$ is the pressure of the ICM around each cavity, $V$ is the cavity volume and t$_{\text{age}}$ is the age of the cavity system. We assume that the 3D shape of each cavity is that of a prolate ellipsoid and compute their volume as $V = (4\pi/3)R_{\text{m}}^{2}R_{\text{M}}$. We combine the temperature of the innermost annulus of Tab. \ref{tab:radialdeproj} with the high-resolution density profile of the ICM to compute the pressure around D1 ($P_{\text{ICM,D1}} = 1.7^{+0.3}_{-0.3}\times 10^{-10}$ erg cm$^{-3}$) and around D2 ($P_{\text{ICM,D1}} = 1.4^{+0.2}_{-0.2}\times 10^{-10}$ erg cm$^{-3}$). 
\\ \noindent The ages of the two cavities have been derived following two approaches (see e.g., \citealt{2012AdAst2012E...6G}):
\begin{enumerate}
	\item We suppose that D1 and D2 are moving at the sound speed $c_{\text{s}}=\sqrt{\gamma kT/\mu m_{\text{p}}}$ of the ICM. Since both the two cavities are inside the innermost annulus used for the deprojected spectral analysis (see Fig. \ref{fig:a_chip}, panel \textit{b}), we use the temperature measured within $r = 16.7$'' ($40.6$ kpc) from the X-ray peak to compute the sound speed: we find $c_{\text{s}} = 938 \pm 43$ km s$^{-1}$. The resulting cavity ages (t$_{\text{age}} = D/c_{\text{s}}$, where $D$ is the projected distance of the cavity from the BCG) are $42.4\pm2.0$ Myr for D1 and $28.8\pm1.3$ Myr for D2.
	\item If the cavity is rising buoyantly in the cluster's atmosphere, its age can be computed as t$_{\text{age}} = D/\sqrt{2gV/SC}$, where g is the gravitational acceleration at the cavity position (measured in Subsect. \ref{hydrostaticmass}), S is the cavity area and C=0.75 is the drag coefficient. We obtain t$_{\text{age,D1}}=83.2\pm12.5$ Myr and t$_{\text{age,D2}}=53.7\pm8.1$ Myr. 
\end{enumerate}
\noindent The difference between the two methods reflects the uncertainty related to the ages of the depressions; therefore, for the following computation we use the average between the two estimate as our best guess for t$_{\text{age}}$, that is $62.8\pm20.4$ Myr for D1 and $41.3\pm12.5$ Myr for D2. As a consequence, the cavity power is $(1.0\pm0.6)\times10^{43}$erg s$^{-1}$ for D1 and $(1.2\pm0.5)\times10^{43}$erg s$^{-1}$ for D2. We summarize the properties of the two depressions in Tab. \ref{tab:cavprop}. 
\begin{table}
\centering
		\caption{Properties of the putative cavities D1 and D2: (1) name of the depression; (2) semi-major axis; (3) semi-minor axis; (4) distance from the BCG; (5) work done to create the cavity; (6) ages computed with the sound speed (first entry) and the buoyant time (second entry); (7) cavity power P$_{\text{cav}}=pV$/$<$t$_{\text{age}}>$.}
		\label{tab:cavprop}
		\begin{tabular}{lcccccc}
			\hline
			& R$_{\text{M}}$& R$_{\text{m}}$ & D  & $pV$  & t$_{\text{age}}$ &  P$_{\text{cav}}$ \\
			& kpc& kpc & kpc  &  10$^{58}$erg &Myr &  10$^{43}$erg s$^{-1}$\\
			\hline
			 D1& 6.4 & 6.2 & 40.6 &  2.0$\pm 0.5$ & \begin{tabular}{@{}c@{}} 42.4$\pm 2.0$ \\ 83.2$\pm$12.5 \end{tabular}&  1.0$\pm 0.6$ \\ 
			 \hline
			 D2& 9.9 & 4.6 & 27.7 &  1.4$\pm 0.2$& \begin{tabular}{@{}c@{}} 28.8$\pm 1.3$ \\ 53.7$\pm$8.1 \end{tabular} &  1.2$\pm 0.5$ \\ 
			\hline
		\end{tabular}
\end{table}	
\\ A strong evidence for the self-regulated feedback loop is that comparisons between the cavity power, $P_{\text{cav}}$, and the cooling luminosity of the ICM, $L_{\text{cool}}$, show that $P_{\text{cav}}$ roughly scales in proportion to $L_{\text{cool}}$ (e.g., \citealt{2007ARA&A..45..117M,2012NJPh...14e5023M}). It is then interesting to compare the total cavity power of D1 and D2 ($P_{\text{cav}} = 2.2\pm0.8\times10^{43}$ erg s$^{-1}$) with the X-ray luminosity of the cooling region by adding the measurements for A795 to those of other clusters (samples of \citealt{2006ApJ...652..216R,2015ApJ...805...35H}). 
\begin{figure}
	\centering
	\includegraphics[width=\hsize]{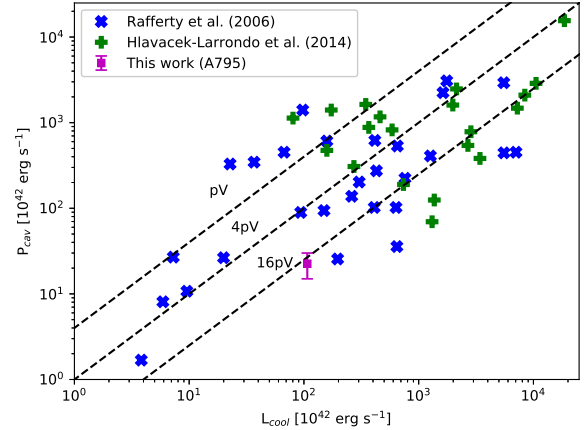}
	\caption{Cavity power vs X-ray cooling luminosity of the ICM. Different symbols denote systems in different samples; the magenta square represents the cavity system of A795, with the error bars expressing the uncertainty in cavity power due to the different age estimates.}
	\label{fig:cavnostre}
\end{figure}
As shown in Fig. \ref{fig:cavnostre}, the putative system of cavities in A795 follows the distribution of the other galaxy cluster samples, indicating that if the two depressions are real cavities, their power is sufficient to offset radiative cooling. 
\linebreak
\\ We point out that D1 and D2 are not symmetric with respect to the AGN: the radio lobes inflated by bipolar jets are expected to be approximately on opposite sides of the nucleus. As AGN-inflated radio bubbles are easily subject to gas motions and turbulence in the central regions of galaxy clusters (e.g., \citealt{2010MNRAS.407.1277M}), we could speculate (similarly to \citealt{2019ApJ...885..111P} for the older generation of cavities in A2495) that sloshing motions of the ICM have influenced the cavity system's direction of motion, leading to the observed asymmetry. \\In fact, the position of D1 and D2 agrees with the direction of motion of the ICM nearby the center (towards north-east, see Fig. \ref{fig:cavpos}). Additionally, the distance of the two cavities from J092405.30+14 is compatible with sloshing motions: we assume that the ICM sloshes with a Mach number in the range $v_{\text{slosh}}^{2}/c_{\text{s}}^{2}=$0.3-0.5 (e.g., \citealt{2006ApJ...650..102A}), and proceed to compute the expected distance of D1 and D2 from the AGN as $v_{\text{slosh}}\times$t$_{\text{age}}$; we obtain $\sim$33-43 kpc for D1 and $\sim$22-28 kpc for D2 (the range expresses the uncertainty in Mach number), to compare with a measured distance of 40.6 kpc for D1 and 27.7 kpc for D2. Far from establishing a causal link between the position of the depressions and the turbulence of the ICM, our result might at least suggest that a sloshing-influenced motion of the cavities is not ruled out. However, the detection of multiple generations of cavities (with deeper observations) is necessary to further explore this hypothesis. 
\\ We note that it is unclear whether the jets of FR0s are able to expand into the external medium and excavate cavities in the ICM. The FIRST observation of J092405.30+14 did not detect 1.4 GHz radio emission corresponding to the two depressions. The flux difference between the NVSS (114$\pm$3.4 mJy, with a resolution of 45''$\sim$110 kpc) and the FIRST (108.25$\pm$0.15 mJy, with a resolution of 5''$\sim$12 kpc) observations at 1.4 GHz of A795 is 5.7$\pm$3.4 mJy: this could be indicative of the presence of extended components, but given the low angular resolution of the NVSS we do not have any information on their morphology (radio lobes or diffuse emission, see Subsect. \ref{diffusemission}).
\\ It is also possible that the structures are filled with an ageing electron population, emitting at lower frequencies. In fact, \citet{2020A&A...642A.107C} observed three FR0s at 150 MHz which show a head-tail structure extending for $\sim$50 kpc, although none of these sources was classified as extended in the FIRST images at 1.4 GHz. Therefore, at MHz frequencies FR0s could reach sizes comparable to the distance between J092405.30+14 and the two depressions we detected (the average distance of D1 and D2 from the AGN is $\sim$34 kpc). 
\\ As a sanity check, we verify whether the 5 GHz luminosity of J092405.30+14 ($L_{\text{5\,\text{GHz}}}$ = 4.52$\pm0.05\times10^{40}$ erg s$^{-1}$, \citealt{2010MNRAS.408.2261K}) could be compatible with the measured cavity power. By studying a sample of sub-Eddington AGNs, \citet{2007MNRAS.381..589M} derived the following relations:
\begin{align}
\log P_{\text{cav}} &= (0.54 \pm 0.09)\log L_{o,5\,\text{GHz}} + 22.1^{+3.5}_{-3.5} \label{pjetlo5}\\
\log P_{\text{cav}} &= (0.81 \pm 0.11)\log L_{i,5\,\text{GHz}} + 11.9^{+4.1}_{-4.4} \label{pjetli5}
\end{align} 
\noindent where $P_{\text{cav}}$ has been estimated from the $pdV$ work done to inflate X-ray cavities and by modelling the radio emission of jets, while $L_{o,5\,\text{GHz}}$ and $L_{i,5\,\text{GHz}}$ are the \textit{observed} (not corrected for beaming) and \textit{intrisic} 5 GHz luminosities, respectively. As we have no information on the beaming factor of J092405.30+14, we used both equations to compute $P_{\text{cav}}$; from Eq. \ref{pjetlo5} we find $P_{\text{cav}}=(1.1\pm0.5)\times10^{44}$ erg s$^{-1}$. We derived the intrinsic 5 GHz luminosity with the \textit{fundamental plane of black hole activity} \citep{2003MNRAS.345.1057M}, obtaining $L_{i,5\,\text{GHz}} = (1.6 \pm 0.8)\times10^{39}$ erg s$^{-1}$; Eq. \ref{pjetli5} gives $P_{\text{cav}}= (4.5\pm2.3)\times10^{43}$ erg s$^{-1}$. Both estimates are $\gtrapprox$ than the measured power $P_{\text{cav}} = 2.2\pm0.8 \times 10^{43}$ erg s$^{-1}$ of D1 and D2.
\\ Furthermore, we check whether the cavity system of A795 follows the trends  $P_{\text{cav}}$-$L_{\text{radio}}$ (total radio power between 10 MHz - 10 GHz) and $P_{\text{cav}}$-$P_{\text{1.4GHz}}$ (monochromatic radio power at 1.4 GHz) reported in \citet{2011ApJ...735...11O}. We use the NVSS observation of A795 (114$\pm$3.4 mJy at 1.4 GHz) to estimate the monochromatic and total radio power associated to the AGN: assuming a spectral index $\alpha=1$ we obtain $L_{\text{radio}}$=1.1$\pm0.1\times10^{41}$ erg s$^{-1}$ and $P_{\text{1.4GHz}}$=5.7$\pm0.2\times10^{24}$ W Hz$^{-1}$. With a cavity power $P_{\text{cav}} = 2.2\pm0.8 \times 10^{43}$ erg s$^{-1}$ we find that the cavity system of A795 lies within the distributions of cavity systems of \citet{2011ApJ...735...11O}. \\
Given the above results, we conclude that the scenario where J092405.30+14 has a sufficient power to inflate the two depressions is energetically consistent. 
\subsection{Sloshing dynamics and heating}
\label{sloshingtime}
\begin{figure}
	\centering
	\includegraphics[width=\hsize]{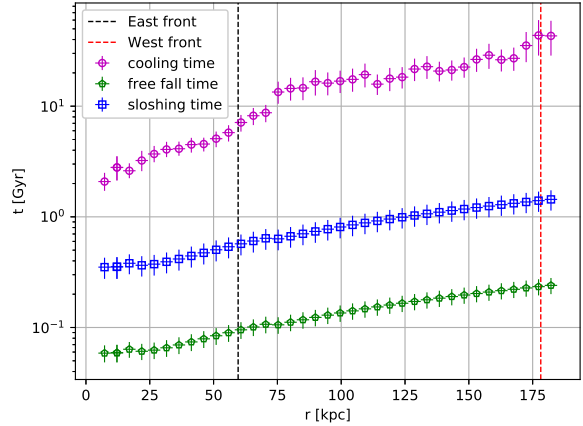}
	\caption{Time-scales radial profiles for A795; the distance of each cold front from the cluster center is marked with a vertical dashed line. As $t_{\text{fall}}$ and $t_{\text{slosh}}$ depend on $M(r)$, which has been computed for $r>$2''$\sim4.86$ kpc (see Fig. \ref{fig:idromass}), we maintain the same radial range for this plot.}
	\label{fig:sloshprofile}
\end{figure}

The analysis of A795 has revealed a cold X-ray gas spiral wrapping around the cluster center; we aim to estimate the \textit{sloshing age} (i.e. the time that has passed since the perturbation has been set) from the thermodynamical analysis we performed on the ICM. To obtain a first, lower limit estimate on the sloshing timescale we compute the free fall time $t_{\text{fall}}(r)$, defined as: 
\begin{equation}
t_{\text{fall}}(r) = \sqrt{\frac{2r^{3}}{GM(r)}}
\end{equation}
\noindent where $r$ is the distance from the cluster center, and $M(r)$ is the total mass profile. 
\\ \noindent This can be seen as the minimum time that is needed for the displaced gas at distance $r$ from the X-ray peak to return to the center, if there was no outward pressure or other forces to counteract gravity. We produce the $t_{\text{fall}}$ profile of Fig. \ref{fig:sloshprofile} (green data), which indicates that the ICM at different radii would take $\sim$50 Myr to $\sim$250 Myr to collapse to the center. However, we expect the true time scale that regulates the sloshing motion to be higher than the free fall time, as the entropy gradient is likely to induce an oscillation of the gas around its equilibrium position, and not a direct collapse to the center \citep{2010A&A...516A..32G}. 
\\ \noindent Indeed, several studies of the sloshing age have considered that the motion of the gas around the cluster center can be approximated as an oscillating flow in a static, stable environment (e.g., \citealt{2003ApJ...590..225C,2017ApJ...851...69S,2020MNRAS.496.1471K}); considering this, the approach to obtain $t_{slosh}$ consists in calculating the Brunt-V{\"a}is{\"a}l{\"a} frequency $\omega_{\text{BV}}$ at each radius r \citep{1990ApJ...357..353B}:
\begin{equation}
\omega_{\text{BV}}(r) = \sqrt{\frac{3GM(r)}{5r^{3}}\frac{d\ln K}{d\ln r}}
\label{omegabv}
\end{equation}
\noindent where $K=kT/n_{\text{e}}^{2/3}$ is the gas entropy, and $M(r)$ is the total mass within $r$; the sloshing timescale is then given by $t_{\text{slosh}} = 2\pi/\omega_{\text{BV}}$. 
\\ We use the hydrostatic mass profile computed in Subsect. \ref{hydrostaticmass} for $M(r)$ and the deprojected entropy profile of the ICM (Fig. \ref{fig:densentpres} panel \textit{c}) to obtain the entropy gradient $d\ln K/d\ln r$. The resulting sloshing timescale profile $t_{\text{slosh}}(r) = 2\pi/\omega_{\text{BV}} (r)$ is presented in Fig. \ref{fig:sloshprofile}. 
We note that $t_{\text{slosh}}(r)$ ranges between 0.4-1.4 Gyr, and raises going far from the center; by considering the distance of the two cold fronts, we can infer the time that has passed since the creation of each discontinuity: 
\begin{enumerate}
	\item The inner, east cold front is at a distance of $59.6 \pm 0.3$ kpc from the cluster center. This suggests a sloshing time of $t_{\text{slosh}}\approx0.57\pm0.12$ Gyr.
	\item The outer, west cold front is situated at $178.2 \pm 2.2$ kpc from the X-ray peak. The resulting age is $t_{\text{slosh}}\approx1.41\pm0.29$ Gyr.
\end{enumerate}
\noindent As expected, the sloshing time scale computed with the Brunt-V{\"a}is{\"a}l{\"a} frequency $\omega_{\text{BV}}$ is higher than the free fall time, specifically by a factor of $\approx 5-6$ (\citealt{2017ApJ...851...69S} found a similar difference between the two estimates in the Fornax cluster). 
\\ To test whether our estimate is reasonable, we compare our results with the simulations of sloshing by \citet{2011MNRAS.413.2057R}: the ages of the simulated cold fronts at $\sim 60$ kpc and $\sim 140$ kpc are approximately 0.7 Gyr and 2 Gyr; these distances and time-scales are comparable to those of our east and west cold fronts. Furthermore, simulations of sloshing have found that cold fronts typically develop after $\approx$1 Gyr from the start of the perturbation (e.g., \citealt{2011MNRAS.413.2057R,2013ApJ...762...69Z,2016JPlPh..82c5301Z}), a value comparable the age of our cold fronts. 
\\We conclude that the event which has offset the ICM from its equilibrium configuration has occurred more than $\approx$1 Gyr ago, and that sloshing is responsible for the formation of the two cold fronts in A795.
\\ Moreover, it has been suggested (e.g., \citealt{2003ApJ...590..225C,2001ApJ...562L.153M}) that as the gas sloshes, the mechanical, turbulent energy of the moving fronts might be converted into heat: we aim at understanding whether sloshing in A795 could represent an additional mechanism - besides AGN heating - capable of offsetting radiative cooling. Fig. \ref{fig:sloshprofile} shows that the sloshing time scale is $\approx 3-13$ times shorter than the cooling time, implying that if the turbulence induced by sloshing in A795 is actually being converted into thermal energy, the process is efficient. \\ We can provide a rough guess of the sloshing heating rate as the kinetic energy $E_{\text{k}}$ associated to each cold front divided by their ages. We assume that the gas masses of the two cold fronts (computed within the innermost spectral bins of Fig. \ref{fig:sectors}) are moving with a Mach number 0.3-0.5 (e.g., \citealt{2006ApJ...650..102A}). We obtain $E_{\text{k,east}} \sim 0.5-0.9\times10^{60}$ erg and $E_{\text{k,west}} \sim 1.5-2.4\times10^{60}$ erg (comparable to the estimates of e.g., \citealt{2013A&A...556A..44R} for A2142 and \citealt{2019ApJ...871..207U} for A907). Considering the estimated ages of the two cold fronts we measure a total sloshing heating rate of $0.6-1.0\times10^{44}$ erg s$^{-1}$, which almost matches the cooling luminosity ($\sim1.1\times10^{44}$ erg s$^{-1}$) derived in Sect. \ref{coolingprop}. As our estimate for the heating rate is likely an upper limit (it is possible that $E_{\text{k}}$ is not entirely converted into heat), we speculate that sloshing in A795 might at least be partially responsible for the reduced cooling of the ICM. 

\subsection{The candidate radio mini-halo}
\label{diffusemission}
\begin{figure}
	\centering
	\includegraphics[width=\hsize]{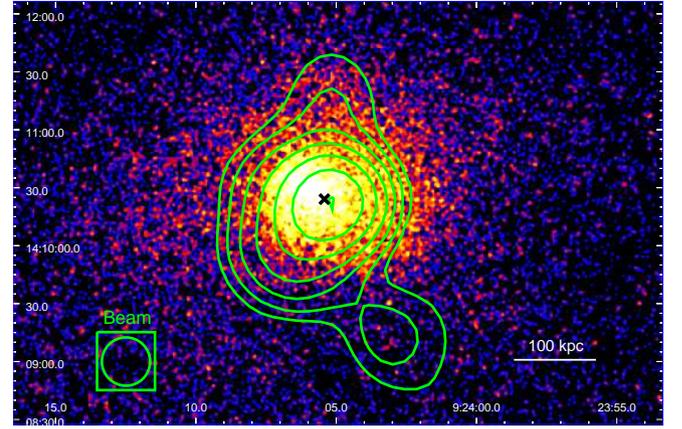}
	\caption{0.5-7 keV images of A795, Gaussian-smoothed with kernel radius of 1.5''. The green contours at 3, 6, 12, 24, 48, 96, 192 $\times\sigma$ are from the TGSS at 150 MHz. The black cross marks the position of the BCG.} 
	\label{fig:gmrtvlss}
\end{figure}

\begin{figure}
	\centering
	\includegraphics[width=\hsize]{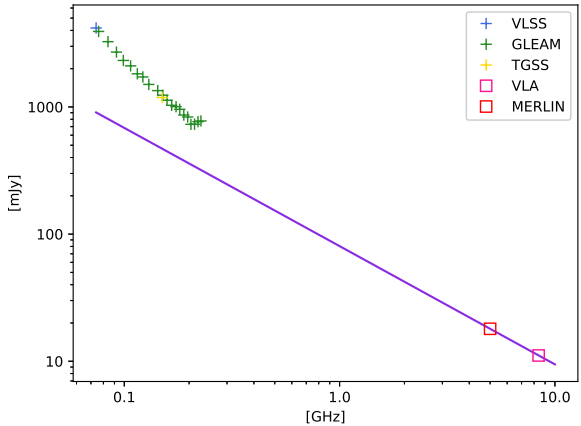}
	\caption{Radio spectrum of A795: the purple line is the extrapolation of the AGN emission at 5 and 8.4 GHz to MHz frequencies; the '+' data are the flux measurements of the VLSS (blue), TGSS (gold) and GLEAM (green) surveys, while the boxes represent the high frequency emission of J092405.30+14.}
	\label{fig:radioA795}
\end{figure}

\begin{figure}
	\centering
	\includegraphics[width=\hsize]{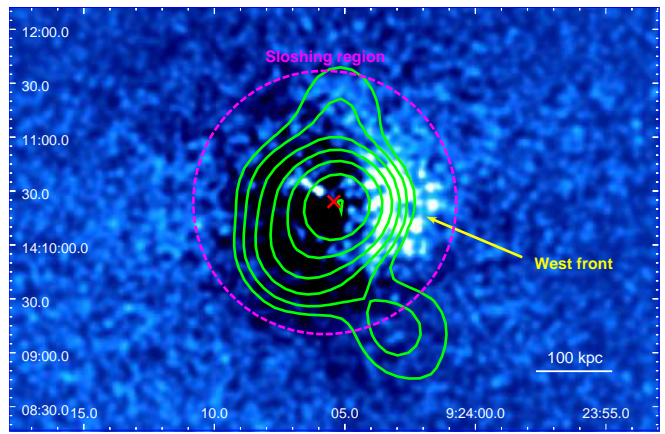}
	\caption{X-ray residual image of A795, with TGSS radio contours (green) as in Fig. \ref{fig:gmrtvlss}. The dashed magenta circle indicates the extension of the sloshing region, defined as the circle with radius equal to the distance of the outermost cold front from the cluster center. The red cross indicates the position of the BCG.}
	\label{fig:residmini}
\end{figure}

To complement our X-ray analysis of A795 and J092405.30+14, we investigate low frequency radio observations of A795 which could provide additional information on the properties of the central AGN. In particular, we find evidence for the presence of extended radio emission over the cluster size in the data archive of the low frequency radio surveys TGSS\footnote{TIFR GMRT sky survey at 150 MHz \citep{2017A&A...598A..78I}.} (150 MHz, resolution of 25''), GLEAM\footnote{GaLactic and Extragalactic All-sky Murchison Widefield Array survey across 72-231 MHz \citep{2017MNRAS.464.1146H}.} (72-231 MHz, resolution of 2'), and VLSS\footnote{VLA Low-frequency Sky Survey at 74 MHz \citep{2007AJ....134.1245C}.} (74 MHz, resolution of 80''). Hereafter we present our interpretation on the radio source nature, based on a morphological and spectral study of the low frequency emission.
\\ The TGSS data at 150 MHz, with a resolution of 25''$\times$25'' and a median noise of $\sigma$ $\sim$3.5 mJy/beam, represent the highest resolution, low frequency radio observation of A795: the radio contours shown in Fig. \ref{fig:gmrtvlss} reveal that the emission is centered on the BCG, and has a largest linear size of 177'' (429 kpc, measured from the 3$\sigma$ contours). Both the VLSS and the GLEAM survey detected extended radio emission coincident with the TGSS contours, but the lower spatial resolution results in a worse characterization of the sub-components; therefore, we use the TGSS for the morphological analysis of the extended emission.
The 150 MHz radio contours show that the extension is significant, with a roundish shape at the center and two smaller protrusions in the north and south-west directions. The average size of the radio source (estimated from the 3$\sigma$ contours at 150 MHz) is $R = $66.7'' ($\sim$162 kpc). 
\\In Fig. \ref{fig:radioA795} we present the spectrum of the radio data for A795. Measuring the fluxes and spectral index associated to the extended emission could provide useful hints at its nature; to accurately separate the emission of the extended source from that of the AGN in the BCG, we proceed with the following steps:
\begin{enumerate}
	\item Considering a single power-law spectrum, we calculate the spectral index of the emission of J092405.30+14 between 5 GHz and 8.4 GHz (see Tab. \ref{tab:minispec}): since at these two frequencies the AGN has been resolved, we do not expect a contribution from the extended source to the measured fluxes. 
	\item We extrapolate the AGN emission to the lowest and highest frequency of the GLEAM, which measured $3925.4 \pm 105.1$ mJy at 76 MHz and $776.3 \pm 38.7$ mJy at 227 MHz: even though the extended emission is poorly resolved by the GLEAM survey, the flux measurements at different frequencies have been performed using the same extraction region and $u-v$ plane coverage; after the extrapolation, we derive the residual fluxes.
	\item We use the residual fluxes to compute the low frequency spectral index of the diffuse radio emission. 
\end{enumerate}
\noindent The results are reported in Tab. \ref{tab:minispec}, while the AGN flux extrapolation to low frequencies is plotted with a purple line in Fig. \ref{fig:radioA795}. \\We note that the low frequency fluxes exceed those expected from the AGN in the BCG. This suggests that the low frequency structure is not directly related to the emission of J092405.30+14, but appears as a diffuse radio source, possibly associated to the ICM, surrounding the AGN. In particular, the power of the diffuse source is $P(227\,\text{MHz})$ = $22.5\pm4.3 \times 10^{24}$ W/Hz and $P(76\,\text{MHz})$ = $152.7\pm15.7\times 10^{24}$ W/Hz. 
\\ \noindent The core spectral index is $\alpha_{\text{c}} = 0.93 \pm 0.09$, and the diffuse emission spectral index is $\alpha_{76-227\,\text{MHz}} = 1.73 \pm 0.86$ (the large uncertainty is due to the relatively large errors of survey flux measurements).
\\We note that these fluxes and powers might suffer a slight contribution from extended, unresolved components of the radio galaxy. In fact, the unresolved VLA observation at 4.8 GHz measured a flux of $24.2\pm0.2$ mJy \citep{2015MNRAS.453.1201H}: the difference of $\approx6$ mJy between VLA and MERLIN might arise from small ($\lessapprox$8.3 kpc), extended features of the central FR0. With the resolutions of the available observations we are unable to remove this contribution from the diffuse source, therefore the residual powers have to be treated as upper limits. Future high angular resolution, low frequency observations of A795 could help us isolating the contribution of any extended component of the radio galaxy from the emission of the diffuse source.
\begin{table*}
	    \centering
		\caption{Radio spectral properties of J092405.30+14 and of the candidate mini-halo: (1) 5 GHz flux of J092405.30+14; (2) 8.4 GHz flux of J092405.30+14; (3) spectral index between 5-8 GHz of J092405.30+14; (4) 227 MHz flux of the diffuse source; (5) 227 MHz power of the diffuse source; (6) 76 MHz flux of the diffuse source; (7) 76 MHz power of the diffuse source; (8) spectral index of the diffuse source between 76-227 MHz.}
		\label{tab:minispec}
		\begin{tabular}{cccccccc}
			\hline
			$S_{\text{5}}$& $S_{\text{8.4}}$ & $\alpha_{\text{c}}$ & $S_{\text{227}}$ & $P_{\text{227}}$ & $S_{\text{76}}$& $P_{\text{76}}$& $\alpha_{\text{76-227 MHz}}$ \\
			mJy&mJy&&mJy&$10^{24}$ W/Hz&mJy&$10^{24}$ W/Hz&\\
			\hline
			\rule{0pt}{2.5ex}18.0$^{+0.2}_{-0.2}$ & 11.1$^{+0.1}_{-0.1}$ & $0.93^{+0.09}_{-0.09}$& $455.2^{+124.4}_{-124.4}$& 22.5$^{+4.3}_{-4.3}$ & $3035.2 ^{+432.2}_{-432.2} $ & 152.7$^{+15.7}_{-15.7}$& $1.73^{+0.86}_{-0.86}$\\
			\hline
		\end{tabular}
\end{table*}
\\ \noindent We also discuss the possibility that the two protrusions on the north and south-east side might not belong to the diffuse source: given their roundish shape and their position at approximately opposite sides of the cluster center, we considered that they could be radio lobes produced by a past activity of the AGN in the BCG. In order to be detected only at MHz frequencies, the radio plasma would have to be very old (see e.g., the 160 Myr old radio lobes found in A449 by \citealt{2016ApJ...817L...1H}), possibly with an ultra-steep spectrum ($\alpha \gtrapprox$ 2, see e.g., \citealt{2019SSRv..215...16V}). The evolutionary properties of FR0s have not been fully explored yet, but it seems unlikely that they experienced a very old radio activity that was able to produce extended emission on scales of a few $\sim$100 kpc (see e.g., \citealt{2019A&A...631A.176C,2019ApJ...871..259G}). Moreover, a past activity of the AGN would have likely produced X-ray cavities coincident with the two extensions, but the two bubbles (especially the northern one, see Fig. \ref{fig:residmini}) do not coincide with any depression in surface brightness. However, as large and external cavities are difficult to detect (see e.g., \citealt{2009AIPC.1201..301B}), it is possible that the \textit{Chandra} observation of A795 is not deep enough to detect surface brightness depressions at the position of the two bubbles. 
\\ In addition to this, the two putative cavities (D1 and D2) in A795 are located north-east of the radio galaxy, which is also the direction of the jet observed at 5 GHz (Fig. 2 in \citealt{2010MNRAS.408.2261K}); on the contrary, the two protrusions are headed west: while a change in the jet direction is possible (see e.g., \citealt{2002MNRAS.330..609D,2018ApJ...852...47R}), it would complicate the radio lobes scenario even further.\\ It may also be that the radio emission arises from unresolved (or barely-resolved) background sources, such as radio galaxies. However, the literature search for objects with coordinates matching those of the two lobes returned no evidences for the presence of radio galaxies. We conclude that it is unlikely that the two extensions have originated from a past radio activity of the BCG, and they are part of the diffuse emission.
\\Considering the above findings, we propose that the extended radio emission in A795 is a candidate mini-halo, possibly powered by the sloshing-induced turbulence of the ICM: the average size of the diffuse source ($\sim$162 kpc) is consistent with the typical sizes of mini-halo, $\approx 50 - 200$ kpc (e.g., \citealt{2008A&A...486L..31C,2019ApJ...880...70G}), and the roundish shape of the TGSS contours resembles the shape of mini-halos. 
\\ It has been proposed that turbulent motion in the CC of galaxy clusters might re-accelerate the radio-emitting electrons and power mini-halos (e.g., \citealt{2002A&A...386..456G,2013ApJ...762...78Z,2016arXiv160300368B}); in this picture, the sloshing motion of the ICM might be a reasonable source of re-acceleration: in sloshing clusters with a mini-halo, the synchrotron emission appears co-spatial with the sloshing region, and confined within the detected cold fronts (e.g., \citealt{2019ApJ...880...70G}). In Fig. \ref{fig:residmini} we over-plot the TGSS radio contours on the $\beta$-model residual image of A795: the radio emission is approximately co-spatial with the sloshing region; the contours nicely follow the shape of the west front, and on the east side the mini-halo appears to be confined within the negative residual spiral. 
\\ In addition to the morphology of the low frequency emission, its spectral properties corroborate our hypothesis: the power of the diffuse source (reported in Tab. \ref{tab:minispec}) is in good agreement with typical mini-halos low frequency powers (e.g., \citealt{2014ApJ...786L..17V,2020MNRAS.499.2934R}), and our estimate for the spectral index ($\sim$1.7) is consistent with typical mini-halos spectral indices (e.g., \citealt{2019ApJ...880...70G}) - albeit a bit steeper. 
\\ We then conclude that the morphological and spectral properties of the extended emission in A795 are consistent with the characteristics of radio mini-halos in CC clusters. 
\section{Summary and Conclusions}
\label{conclusion}
This work unveiled the X-ray properties of A795, a weakly CC galaxy cluster displaying signs of dynamical disturbances, and provided a first inspection of the relation between a FR0 radio galaxy and the surrounding ICM. Here we summarize our main results. 
\begin{enumerate}
	\item We determined the global properties of the ICM in A795: within 405 kpc from the cluster center, we measured a temperature $kT = 4.6\pm0.1$ keV, a metallicity of $0.38\pm0.05$ Z$_{\odot}$, and a luminosity $L(0.5-7\,\text{keV}) = 3.43^{+0.04}_{-0.03}\times10^{44}$ erg s$^{-1}$. From the spatially-resolved spectral analysis of the ICM we deduced that A795 is a weakly CC cluster, with a cooling time $<7.7\,\text{Gyr}$ inside $r_{\text{cool}} = 66 \pm 3$ kpc. 
	\item We discovered a cool ICM spiral in A795, which indicates that the gas is sloshing: this mechanism is responsible for the formation of the two observed surface brightness discontinuities, and explains the offset between the X-ray peak and the BCG. The spectral study of the two edges has confirmed their cold fronts nature: for the east front we measured a temperature ratio $T_{\text{out}}/T_{\text{in}} = 2.07 \pm 0.53$ and a density ratio of $n_{\text{in}}/n_{\text{out}} = 1.69 \pm 0.07$, while the west front has $T_{\text{out}}/T_{\text{in}} =1.61 \pm 0.46$ and  $n_{\text{in}}/n_{\text{out}} = 1.93 \pm 0.31$; we found pressure equilibrium and an entropy jump at both discontinuities. 
	\item The X-ray emission of J092405.30+14, the FR0 at the center of A795, is typical of radio loud AGN powered by radiatively inefficient, advection dominated accretion flows: we measured a steep photon index $\Gamma = 2.13^{+0.16}_{-0.15}$ for the X-ray emission. By comparing the temperature, density, and dynamics of the surrounding ICM with those of ICM around typical FRIs in BCGs, we concluded that the environment alone cannot explain the observed radio size of this new class of sources, and that an intrinsic jet weakness is likely necessary.
	\item We identified two putative cavities at an average distance of $\approx$34 kpc from the central AGN: by measuring their power ($P_{\text{cav}} = 2.2 \pm 0.8 \times 10^{43}$ erg s$^{-1}$) and comparing it to the bolometric luminosity within the cooling region ($L_{\text{cool}} = (1.07 \pm 0.06)\times 10^{44}$ erg s$^{-1}$), we found that these cavities might be able to offset and reduce the cooling efficiency in A795. 
	\item By computing the timescale over which the gas oscillates at each radius, we inferred that the perturbation of the ICM has been set $\gtrapprox 1$ Gyr ago. Moreover, we speculated that the kinetic energy of the cold fronts might concur the heating of the ICM. 
	\item The 150 MHz archival data for this cluster revealed the presence of extended radio emission: the low frequency flux of this component exceeds the flux extrapolated from the high frequency (5-8 GHz) emission of the central AGN; considering this finding, and the roundish shape of the extended emission (with R$\approx$162 kpc), we classified it as a candidate mini-halo. The mini-halo power is $P_{\text{MH, 227}}\lessapprox22.5\times 10^{24}$ W/Hz at 227 MHz, and $P_{\text{MH, 76}}\lessapprox152.7\times 10^{24}$ W/Hz at 76 MHz. We estimated a low frequency spectral index of the candidate mini-halo of $\alpha_{\text{MH}} = 1.73 \pm 0.86$ between 76 MHz and 227 MHz. It is possible that the electrons responsible for the radio emission have been re-accelerated by the turbulent motion of the ICM, since the diffuse source and the sloshing region are approximately co-spatial.  
\end{enumerate}
\noindent This is the first in-depth, dedicated \textit{Chandra} study of A795 and of the link between this cluster and the central FR0 radio galaxy. Future multi-wavelength studies will surely provide additional clues on the properties of this cluster and on the interaction between the central AGN and the surrounding environment. In particular, deeper X-ray observations of this cluster and possibly tailored simulations will allow to better characterize the sloshing dynamics. Future low frequency, high resolution radio observations of A795 will be also essential to perform a detailed study of the extended source and to explore the MHz emission of the central FR0.
\\ \noindent We note that the radio galaxy J092405.30+14 is not the unique FR0 in a cluster of galaxies. Therefore, a comparison of the central and environmental properties of other FR0s in clusters will be useful to confirm our results and to better characterize the behavior of this new class of radio sources. 

\section*{Acknowledgements}
We thank the referee for her/his constructive and useful comments, which helped us improve our work and deepen specific aspects of the analysis. The scientific results reported in this article are based on data obtained from the \textit{Chandra Data Archive}. 
\section*{Data Availability}
The data used to produce plots are available on request to the authors. \textit{Chandra} data analyzed in this article are available at \url{cxc.cfa.harvard.edu}. The linear fitting library can be found at \url{https://github.com/rsnemmen/BCES}.



\bibliographystyle{mnras}
\bibliography{A795_bib} 

\bsp	
\label{lastpage}
\end{document}